\documentclass[preprint,11pt]{elsarticle}

\usepackage{amssymb}
\usepackage{amsmath}

\journal{}
\usepackage{hyperref}  
\usepackage{subcaption} 
\usepackage{graphicx}
\usepackage{courier}    
\usepackage{longtable}  
\usepackage{array}
\usepackage{xcolor}  
\usepackage{multicol}  
\usepackage{float} 
\usepackage[utf8]{inputenc}
\DeclareUnicodeCharacter{03BC}{\$mu}
\usepackage{enumitem}
\usepackage{tabularx} 
\usepackage{booktabs} 
\usepackage{adjustbox}
\renewcommand{\arraystretch}{1.1}
\usepackage{placeins}

\usepackage{makecell}

\usepackage{multicol}  
\usepackage{tikz} 

\setcellgapes{2pt}\makegapedcells

\DeclareUnicodeCharacter{FFFD}{ }

\begin{document}

\begin{frontmatter}

\title{RADIANT-LLM: an Agentic Retrieval Augmented Generation Framework for Reliable Decision Support in Safety-Critical Nuclear Engineering}

\author[NE]{Zavier Ndum Ndum\corref{cor1}}
\author[PVFA]{Jian Tao}
\author[NE]{John Ford}
\author[NE]{Mansung Yim}
\author[NE]{Yang Liu\corref{cor1}}

\address[NE]{Department of Nuclear Engineering, Texas A\&M University, College Station, TX, USA}
\address[PVFA]{College of Performance, Visualization and Fine Arts, Texas A\&M University, College Station, TX, USA}
\cortext[cor1]{Corresponding authors: zavier.ndum@tamu.edu, y-liu@tamu.edu}


\begin{abstract}

Reliable decision support in nuclear engineering requires traceable, domain-grounded knowledge retrieval, yet safety and risk analysis workflows remain hampered by fragmented documentation and hallucination when use pre-trained large language model (LLM) in specialized nuclear domains. To address these challenges, this paper presents RADIANT-LLM (Retrival-Augumented, Domain-Intelligent Agent for Nuclear Technologies using LLM), a multi-modal retrieval-augmented generation (RAG) framework designed for nuclear safety, security, and safeguards applications. The framework uses a local-first, model-agnostic architecture that pairs a multi-modal document ingestion pipeline with a structured, metadata-rich knowledge base, supporting page- and figure-level retrieval from technical documents. An agentic layer coordinates domain-specific tools, enforces citation-backed responses with provenance tracking, and supports human-in-the-loop validation to reduce hallucination risks.

To rigorously evaluate this framework, we develop and apply a suite of domain-aware metrics, including Context Precision (CoP), Hallucination Rate (HR), and Visual Recall (ViR), to expert-curated benchmarks derived from Used Nuclear Fuel Storage Facility design guidance. Across varying knowledge base sizes, CoP and ViR remain within an 85--98\% band, and hallucination rates are substantially lower than those observed in general-purpose deployments. When the same queries are posed to commercial LLM platforms without the RAG layer, hallucinations and citation errors increase markedly. These results indicate that a locally controlled, multi-modal RAG framework with domain-specific retrieval and provenance enforcement is necessary to achieve the factual accuracy, transparency, and auditability that nuclear engineering workflows demand.

\end{abstract}

\begin{keyword}
Reliable decision support \sep Large language models \sep Visual retrieval-augmented generation \sep Multi-modal knowledge management \sep Safety-critical systems in nuclear engineering

\end{keyword}

\end{frontmatter}



\section{Introduction}
\label{sec:introduction}

Reliability and safety of complex engineered systems depend on the ability to rapidly synthesize information from regulatory documents, operational procedures, incident reports, and technical design guidance~\cite{XIAO2026112123, LI2026112277, ZHANG2026112333}. This is true across high-risk industries---nuclear energy, aviation, maritime transportation, and chemical processing---where incomplete or delayed information retrieval can directly compromise risk analysis, fault diagnosis, and operational decision making. Recent advances in artificial intelligence (AI) and machine learning (ML), particularly generative AI (GenAI), have produced models with strong natural language understanding and automated reasoning capabilities~\cite{Zhang2024ADiscovery,Zhao2023AModels}, creating practical opportunities for decision support in these safety-critical settings. In nuclear engineering, AI/ML methods have already been applied to uncertainty quantification for multiphase CFD and sodium fast reactor thermal stratification modeling~\cite{LIU2021107636, ABULAWI2025111353}, as well as to AI-driven thermal-fluid testbeds that integrate digital twins and LLM-enabled operator assistance~\cite{LIM2025100023}. Large language models (LLMs), such as OpenAI's ChatGPT, Google's Gemini, and Meta's Llama, are rapidly transforming digital workflows by enabling rapid text translation, summarization, and complex question answering~\cite{abouammoh2025perceptions}. However, the Nuclear Science and Engineering sector faces a persistent ``knowledge management'' bottleneck that general-purpose AI tools do not immediately solve. Critical technical data, regulatory guidance, and modeling and simulation (M\&S) benchmarks are often fragmented across disparate international and national silos, including the International Atomic Energy Agency (IAEA), U.S. Nuclear Regulatory Commission (NRC), and the U.S. Office of Scientific and Technical Information. The volume and heterogeneity of these authoritative resources frequently overwhelm professionals, making efficient retrieval difficult---particularly for nuclear Safety, Security, and Safeguards (popularly known as \textit{nuclear-3S}) applications. By automating the retrieval, organization, and cross-referencing of regulatory and design knowledge, and by streamlining key M\&S tasks, LLM-based tools and AI agents can provide decision support for both existing and emerging nuclear technologies. They can help engineers integrate 3S considerations earlier in the design process and maintain 3S-by-design as concepts evolve toward licensing, deployment, and operation.

\subsection{Opportunities for Generative AI in Nuclear Security, Safety and Safeguards}
\label{subsec:domain_aligned_LLMs}

\begin{figure}[h]
    \centering
    \includegraphics[width=0.85\linewidth]{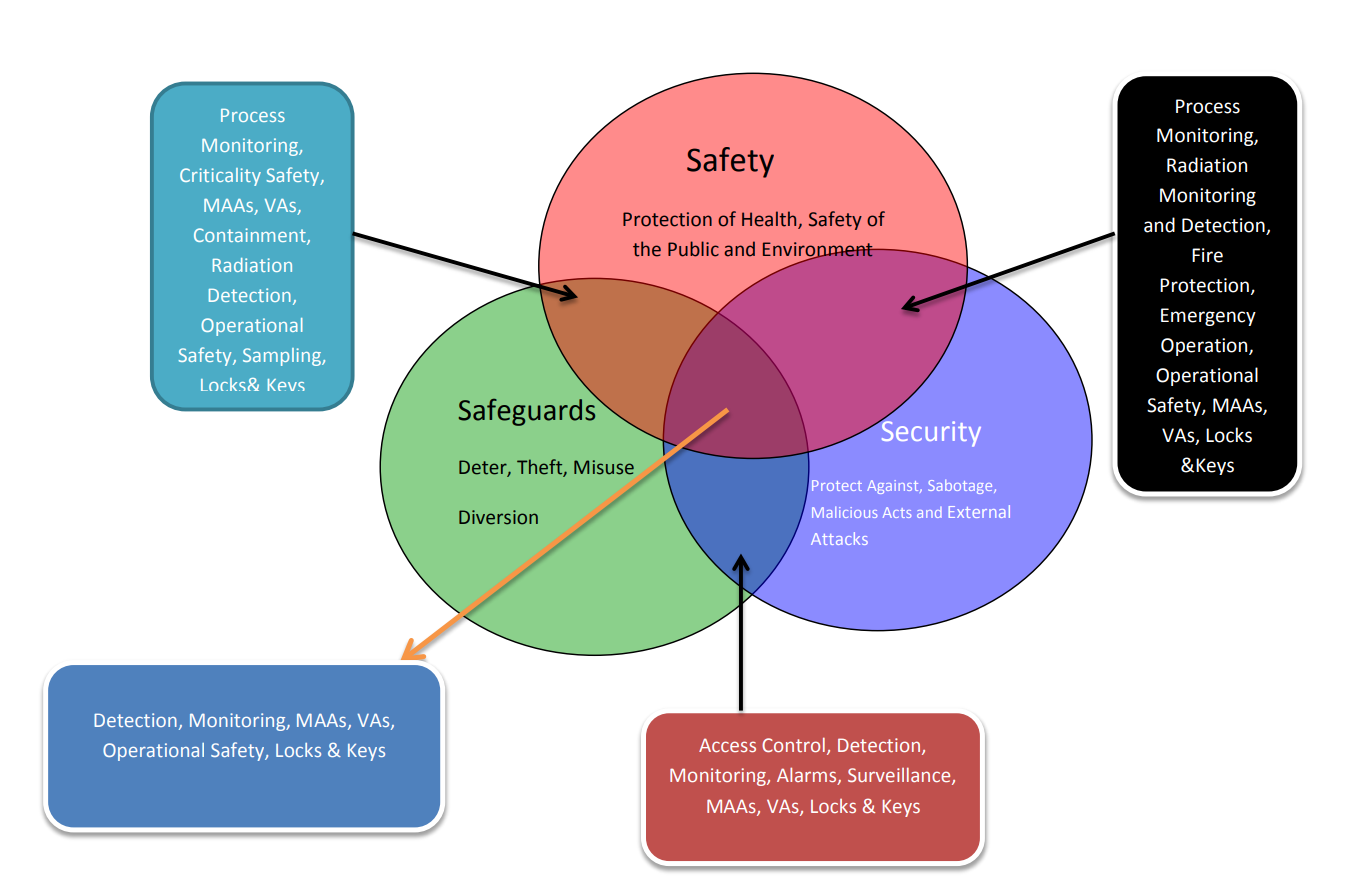} 
    \caption{Interfaces and interdependencies between Safety, Security, and Safeguards (3S) in nuclear facility design. Adapted from Badwan and Demuth~\cite{badwan2015application}.}
    \label{fig:3s_diagram}
\end{figure}

The global energy sector is shifting as nations pursue climate targets while meeting growing electricity demand from digital infrastructure and AI-driven data centers. Nuclear energy, with its proven capacity for reliable, low-carbon power generation, is expected to play a larger role in this transition. The emergence of advanced Small Modular Reactors (SMRs) and Microreactors (MRs) offers potential advantages in safety, construction time and cost, and deployment flexibility~\cite{KOO2025101632}. As defined by the IAEA and illustrated in Figure~\ref{fig:3s_diagram}, \textbf{Safety} focuses on preventing accidents and mitigating radiological consequences; \textbf{Security} addresses the physical and cyber protection of nuclear materials and facilities against malicious acts; and \textbf{Safeguards} comprise technical measures and accounting systems to ensure that nuclear material is not diverted from peaceful uses to weapons programs~\cite{badwan2015application}.

A central engineering challenge is managing interface conflicts among these domains. For example, a safety requirement for rapid emergency egress may conflict with a security requirement for delay barriers. Also, additional surveillance measures to support safeguards may introduce a cyber-attack surface that must be managed under security requirements. Resolving such trade-offs requires engineers to synthesize heterogeneous information, including regulatory guides, IAEA safety standards, physical protection and safeguards guidance, facility schematics (P\&IDs), and operational procedures. In parallel, workflows to evaluate the radiation consequences, thermal hydraulics, and other safety analyses of these advanced nuclear technologies built on decades computer modeling, using codes such as MOOSE, SAM, MCNP, OpenMC, and FLUKA~\cite{godoy2020workflows, turinsky2016modeling, turner2016virtual}. These workflows often produce large volumes of data (input files, scripts, and reports) that must remain traceable to the underlying 3S basis. Manual retrieval and cross-checking across these sources are time-consuming and prone to omission, particularly under time-sensitive aggressive design and licensing schedules.

General-purpose LLMs can assist with text search and summarization but are fundamentally constrained in NSE by three issues: (i) \textit{domain reliability}, (ii) \textit{data security and privacy}, and (iii) \textit{operational cost}. In terms of domain reliability, they could be classified as \textit{jacks of all trades}. This is because they are trained on vast public corpora and exhibit only superficial fluency in nuclear concepts. When pressed for specifics, such as thermal-hydraulic and radiation safety margins, plant-specific standard operating procedures (SOPs), safeguards, or security recommendations, they tend to hallucinate, generating plausible but incorrect or fabricated outputs~\cite{zhang2025citation, peereboom2025cognitive, annepaka2025large}. With respect to data security and privacy, uploading sensitive or proprietary documents to commercial or public chatbots can introduce unacceptable risks of data leakage and theft in a sector that handles intellectual property, design details, sensitive security information, and Export Controlled Information (ECI). Modern LLM platforms, which increasingly function as AI agents capable of \textit{tool orchestration} (web search, file search, code execution), could further increase the attack surface and reduce user control over where data is sent, how web content is filtered, and which external sources are consulted. Even organization-led LLM platforms that channel multiple commercial chatbots may reduce \textit{operational costs}, but they do not fully resolve domain reliability or provide the strict control over data flows and the security guardrails required for sensitive nuclear workflows. These limitations point to a need for domain-aligned, locally controlled AI systems that can operate within well-defined 3S and information-security constraints.

At the same time, the rapid evolution of generative AI has introduced new concerns regarding security, governance, and trustworthiness~\cite{spelda2025security, zaidan2024ai, judge2025code}. Recent efforts on AI governance and risk management highlight the need for transparent, controllable systems that align with existing regulatory and oversight frameworks~\cite{smuha2021race, NIST.AI.800-1.ipd, act2025regulation}. For the nuclear sector, this implies that AI tools should operate over locally controlled knowledge bases (KBs), provide traceable citations, and integrate cleanly with established 3S and information-security requirements, which places strong demands on the underlying document-processing and retrieval pipeline. Advances in scientific PDF parsing, including neural OCR-based models such as Nougat~\cite{blecher2023nougat} and newer layout-aware tools like Marker~\cite{marker2025} and Docling~\cite{auer2024docling}, have improved robustness on complex documents but remain largely focused on the ingestion stage of RAG and often degrade on scanned or visually dense technical material~\cite{zhang2024document}. Our prior work~\cite{ndum2025automating} showed that Nougat substantially improves text and equation extraction yet still misses information encoded in figures and graphs without an additional vision pipeline and separate indexing logic. Generic ``chat-with-PDF'' solutions typically lack rigorous source traceability, struggle with dense or scanned content, and rely on remote, opaque vector stores that raise data-sovereignty concerns in the nuclear domain. These gaps motivate the need for specific, domain-aligned solutions.


Recent work in the reliability and safety engineering literature has begun applying LLMs to safety-critical decision support across several industrial domains. In fault diagnosis, Zheng et al.~\cite{ZHENG2024110382} fine-tuned pre-trained LLMs on high-speed train and chemical plant datasets, finding that both open-source and closed-source models achieved high diagnostic performance while also revealing how dataset size, data normalization, and missing values affect diagnosis explainability. In accident investigation, Zhang et al.~\cite{ZHANG2026112333} combined a hierarchical graph attention network with fine-tuned LLMs and RAG for ship collision analysis, producing outputs consistent with real accident reports in causation and responsibility assignment. Li et al.~\cite{LI2026112277} took a related approach for aviation risk classification, using LLM-based data augmentation and Bayesian network modeling to extract high-order risk propagation paths from narrative incident reports.

Research has established that augmenting LLMs with supplementary or domain-specific knowledge can further enhance their performance. Retrieval-augmented generation (RAG), introduced by Lewis et al.~\cite{lewis2020retrieval}, has emerged as a central GenAI technique for this purpose. In RAG, a retriever first identifies relevant documents or passages from a large corpus, and a generator then produces answers using both the retrieved content and its pretrained knowledge~\cite{fan2024survey, gao2023retrieval}. For infrastructure maintenance, Yu et al.~\cite{YU2026111973} proposed a lightweight multimodal LLM dual-agent framework for tunnel defect detection that separates visual recognition from regulatory-compliant reasoning through a RAG module, with the goal of ensuring adherence to national standards while reducing hallucination risks. In maritime operations, Liu et al.~\cite{LIU2026111891} presented Pirate-GPT, a locally deployed LLM framework for offline anti-piracy decision support that uses semantic segmentation, agent collaboration, and two-stage retrieval to support privacy-preserving deployment in resource-constrained environments. Liang et al.~\cite{LIANG2026112416} integrated domain-specific LLMs with functional resonance analysis and Bayesian networks for battery energy storage risk analysis, aiming to reduce reliance on subjective expert knowledge.

Closer to the nuclear domain, Xiao et al.~\cite{XIAO2026112123} developed AutoGraph, a neuro-symbolic knowledge-graph agent for procedure automation and human reliability support in digitalized control rooms. Applied to high-temperature gas-cooled reactor scenarios, AutoGraph achieved a 90\% performance improvement over manual operation by combining LLM-driven semantic parsing with graph-constrained spatial reasoning under zero-tolerance safety constraints. The same group developed KRAIL~\cite{XIAO2026111585}, a framework for human reliability analysis that integrates IDHEAS-DATA, knowledge graphs, and LLMs. In KRAIL, the knowledge graph acts as a RAG layer to surface context-relevant evidence, while an expert-in-the-loop validation step curbs hallucinations; the system produces more accurate human error probability estimates than prior methods in under 150 seconds.

Taken together, these studies confirm that LLM-based frameworks can improve reliability and safety workflows in multiple sectors, but they also expose recurring difficulties: hallucinations in the absence of domain grounding, limited traceability of model outputs, and data-sovereignty concerns when sensitive operational documents are involved. Moreover, these frameworks are predominantly tailored to single modalities (text or structured data) and specific analytical tasks such as fault diagnosis, risk classification, or human reliability assessment. A multi-modal framework that jointly processes text, figures, schematics, and regulatory documents to support safety-critical decision making in nuclear engineering has not yet been reported.

Within the nuclear sector specifically, several studies have already explored LLM applications. Kwon et al.~\cite{kwon2024sentiment} showed that LLMs outperformed traditional machine learning classifiers in analyzing public sentiment on nuclear topics, achieving up to 96\% classification accuracy. Recent work continues to advance RAG toward high-performance scientific deployments (e.g., HiPerRAG)~\cite{gokdemir2025hiperrag}. By grounding responses in explicit documents, RAG can reduce hallucinations and compensate for outdated or incomplete pretraining. For instance, Iob et al.~\cite{iob2024nuclear} showed that LLM-generated training scenarios for nuclear infrastructure personnel, when provided with appropriate context, aligned more closely with those designed by human experts. Retrieval-augmented LLMs (RA-LLMs) have also enabled the summarization of large-scale experimental data~\cite{suresh2024towards}, assisted detector design~\cite{diefenthaler2024ai}, advanced material discovery~\cite{wang2025dreams}, and supported human reliability analysis for nuclear safety~\cite{xiao2024krail}. In health physics and radiological protection, models such as OpenAI's GPT-4 and GPT-4o have shown promise, but still fall short of the nuanced expertise and factual precision required for critical applications~\cite{roemer2024artificial, oumano2025comparison}.

Despite these promising progresses, a gap remains. To our knowledge, no existing solution delivers a secure, domain-aligned, multi-modal agentic AI framework that supports reliable decision making, knowledge management, and compliance for nuclear professionals while maintaining data security, provenance traceability, and domain specificity across text, figures, and regulatory documents.

\subsection{Contributions of this Work}
\label{subsec:contributions}

To address this gap, we present RADIANT-LLM: a Retrieval-Augmented Domain Intelligent Agentic framework for Nuclear Technologies leveraging LLM. RADIANT-LLM is a secure, \textit{training-free} AI assistant designed to run locally on the user's computer with a custom Visual-RAG pipeline. This pipeline employs a local-first architecture for document processing and storage, converting PDFs and images into a persistent, structured, multi-modal knowledge base (KB). Frozen (pre-trained) LMMs are then connected to this local KB for querying via Application Programming Interfaces (API). These APIs function strictly as: (i) a bridge-facilitating in-context learning and local task execution, and (ii) a firewall-enabling secure, controlled access to the LLM models without exposing critical data, thus preserving privacy.

From a technical standpoint, the main contributions of this work are as follows: (i) we present a secure, locally deployable GenAI framework that enables nuclear organizations to apply LLM-based assistance directly to proprietary design documents and sensitive 3S workflows without uploading raw data to external servers; (ii) we develop an advanced, multi-modal \textbf{Visual-RAG} architecture that combines state-of-the-art PDF parsing with image and figure processing, allowing the system to retrieve and use information contained in text, equations, tables, and visual schematics (figures, graphs and charts); (iii) we introduce a dynamic KB construction method in which RADIANT-LLM can start from a small set of seed documents, parse references, automatically identify and download additional public sources under user supervision, and continuously expand its structured, local JSON-based KB; (iv) we design an agentic layer and toolset to explicitly address nuclear 3S requirements by enforcing traceable, citation-backed responses that include document titles, authors, page numbers, and figure references, while aligning with international recommendations to avoid disclosing confidential information or guidance that could facilitate misuse of nuclear technology; and (v) we demonstrate the application of RADIANT-LLM to KBs covering nuclear safety, security, safeguards, design of used nuclear fuel storage facilities (UNFSF), and AI integration into nuclear KBs, showing that the agent can maintain focused attention on these topics and support reliable, traceable knowledge management.

The Visual-RAG architecture presented in this work builds on a sequence of prior developments for NSE workflows. The first stage was introduced in the \textit{AutoFLUKA} framework~\cite{ndum2025automating}, which implemented a semi-advanced, text-centric RAG architecture to support Monte Carlo (MC) simulation workflows in the FLUKA heavy particle transport code. AutoFLUKA extended baseline RAG by integrating robust PDF parsing tools, including Nougat~\cite{blecher2023nougat} and PyMuPDF, to extract structured text, equations, and tables from FLUKA documentation, and used a JSON-based registry to incrementally build a knowledge base. The second stage, a preliminary RADIANT-LLM study~\cite{Ndum2025RadiantLLM}, introduced metadata-aware retrieval, document provenance tracking, and version gating to restrict retrieval to authoritative and up-to-date sources, thereby improving answer reliability and traceability for nuclear-3S applications. The present work represents the third stage: an end-to-end, agentic Visual-RAG architecture that explicitly integrates visual evidence and tool-calling into the retrieval and reasoning loop for nuclear 3S and broader NSE tasks.

\textit{RADIANT-LLM is not intended to replace nuclear professionals; it is a personalized AI assistant that accelerates labor-intensive, repetitive tasks}. The platform is fully customizable, allowing users to integrate their own knowledge sources and tailor the system to specific workflows, including those involving proprietary Generation IV reactor concepts and documents subject to ECI restrictions. Unlike simple document-upload solutions, RADIANT-LLM follows a local-first architecture in which PDFs, images, and other files are processed and stored on the user’s infrastructure, while state-of-the-art pre-trained LLMs are accessed via secure APIs that act as a \textbf{controlled bridge} for in-context learning and task execution and as a \textbf{firewall} that prevents direct exposure of raw documents and intermediate data. 

Retrieval-augmented LLMs (RA-LLMs) can also introduce risks through the injection of compromised or misleading sources. RADIANT-LLM mitigates this by giving users full control over which documents to populate the KB and by making the origin of retrieved content explicit, thereby supporting transparency and trust in the information pipeline. Security-by-design is a foundational principle: the system is reinforced to avoid releasing confidential information found in documents (for example, personal data, grades, salary details, addresses) and is configured to avoid providing legal advice or detailed guidance that could facilitate diversion or misuse of nuclear technology. These safeguards are implemented in line with international standards and recommendations from bodies such as the IAEA~\cite{taeihagh2025governance, judge2025code, spelda2025security} and the U.S. NRC~\cite{smuha2021race, NIST.AI.800-1.ipd, act2025regulation}, ensuring that RADIANT-LLM supports high standards of safety- and security-by-design.

To evaluate both RADIANT-LLM and general-purpose deployments, we further propose a unified, quantitative benchmarking methodology tailored to nuclear workflows. We introduce a set of domain-aware metrics, namely, \textbf{Context Precision (CoP), Citation Precision (CiP), Citation Hit (CiH), Hallucination Rate (HR)}, and \textbf{Visual Recall (ViR)}, which require decomposing each model-generated answer into a set of atomic claims and scoring these against expert-curated gold standards (ground truths). Page-level sensitivity tests probe how well vision–language models recover schematic details and avoid hallucinations, while corpus-level context-scaling experiments examine how retrieval quality evolves as the KB grows from a few documents to hundreds of sources. Across these benchmarks, RADIANT-LLM’s Visual-RAG configuration maintained a consistently high performance, with CoP and ViR staying within approximately 85--98\% over the full corpus. In contrast, general-purpose, long-context, web-enabled deployments performed markedly worse, particularly in enforcing citations and avoiding plausible but factually incorrect quantitative statements. An occasional dip and immediate recovery, as reported in Section~\ref{subsec:Context_Expansion}, was observed around an intermediate KB size (on the order of $\mathcal{O}(10^2)$ sources). This could be attributable to the stochastic, auto-regressive nature of LLM sampling (elaborated in Section~\ref{sec:background}) of this study. It does not undermine the overall robustness of the Visual-RAG pipeline. Even in this era of large-context, high-reasoning LLMs with web search capabilities, a dedicated, locally controlled multi-modal RAG layer remains necessary to maintain factual fidelity, visual grounding, and citation discipline for nuclear 3S workflows.

The remainder of this paper is organized as follows. Section~\ref{sec:background} introduces the related works. Section~\ref{sec:methods} details the methodology of the visual-RAG and agentic AI pipelines. Section~\ref{sec:results} presents the results, highlighting the system's performance on the nuclear-3S benchmark, and comparison with general-purpose platform deployments. Finally, section~\ref{sec:Conclusion} concludes the paper with future perspectives.


\section{Related Work}
\label{sec:background}


Two common approaches are used to adapt LLMs for domain tasks: (i) \emph{model fine tuning}, which updates network weights using domain data, and (ii) \emph{augmentation}, which keeps a pretrained model frozen and adds domain context at inference time through prompting, retrieval, and tool use. Fine tuning can help on narrow tasks, but it typically requires curated datasets and substantial compute, and it couples a deployment to a specific model checkpoint. In nuclear engineering, where guidance and design assumptions evolve and systems may be maintained for long periods, that coupling is a practical constraint. Augmentation avoids this by moving domain knowledge and controls into prompts, retrieval, and external tools, which also simplifies version updates and provenance tracking.

\begin{figure}[htbp]
    \centering
    \includegraphics[width=1\linewidth]{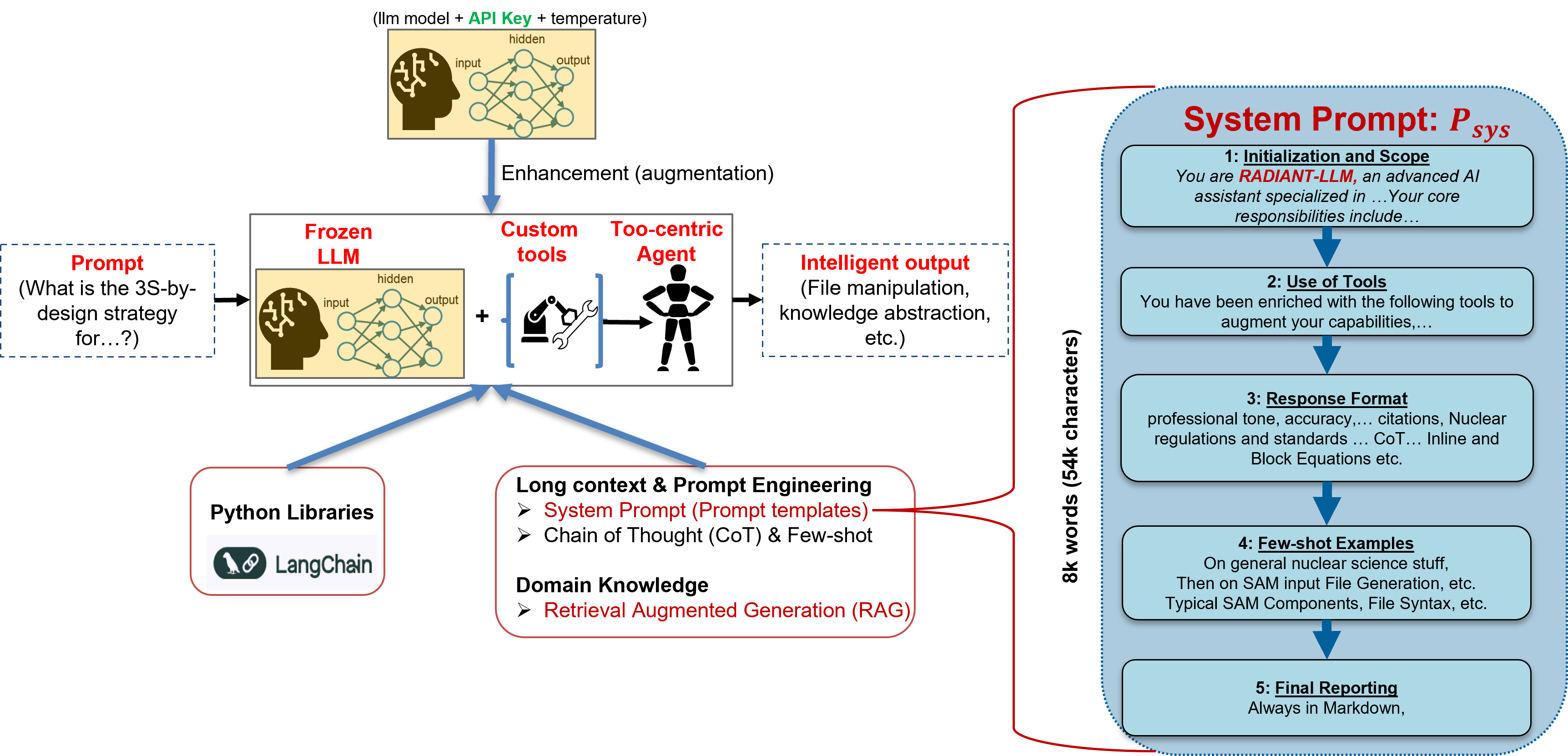}
    \caption{Conceptual illustration of LLM augmentation in RADIANT-LLM. A frozen pretrained language model (yellow) acts as a replaceable reasoning core and is supported at inference time by structured system prompts, RAG, and custom tool calls coordinated by a supervisor agent. This approach enables local deployment, explicit citation requirements, and multimodal reasoning without fine tuning model weights.}
    \label{fig:llm_augmentation}
\end{figure}

This work adopts augmentation rather than fine tuning as the primary enhancement strategy. As illustrated in Figure~\ref{fig:llm_augmentation}, RADIANT-LLM treats the pretrained LLM as a replaceable reasoning core and enforces domain behavior through a modular tool layer. We use two complementary mechanisms. Prompt conditioning leverages long context and structured system instructions, including constrained answer formats and a small number of exemplars, to encourage conservative and self checking responses. Retrieval augmentation couples the LLM to an external document store~\cite{fan2024survey, gao2023retrieval}: a retriever selects passages from a corpus (e.g., PDFs, reports, and standards) and the LLM generates an answer conditioned on both its internal knowledge and the retrieved text. Prior studies report that RAG can improve factual accuracy and reduce hallucinations across domains~\cite{suresh2024towards, diefenthaler2024ai, wang2025dreams}. In our setting, retrieval, prompts, and tools work together to provide practical domain alignment without retraining.

We implemented a small set of domain utilities (engineering checks, file operations, and output schemas) and integrated them into the agent loop using LangChain~\cite{H.Chase2023GitHubApplications}. We selected LangChain because it provides a mature interface for tool orchestration and has been used in a range of LLM applications~\cite{Pandya2023AutomatingOrganizations, Jeong2023GenerativeFramework, Singh2024RevolutionizingModel}. Figure~\ref{fig:llm_augmentation} summarizes the resulting augmentation concept: the frozen model is enhanced at inference time through prompts, RAG, and tool calls, without modifying its parameters.

\section{Methodology: RADIANT-LLM's Agentic Framework}
\label{sec:methods}

\begin{figure}[htbp]
    \centering
    \includegraphics[width=1.0\linewidth]{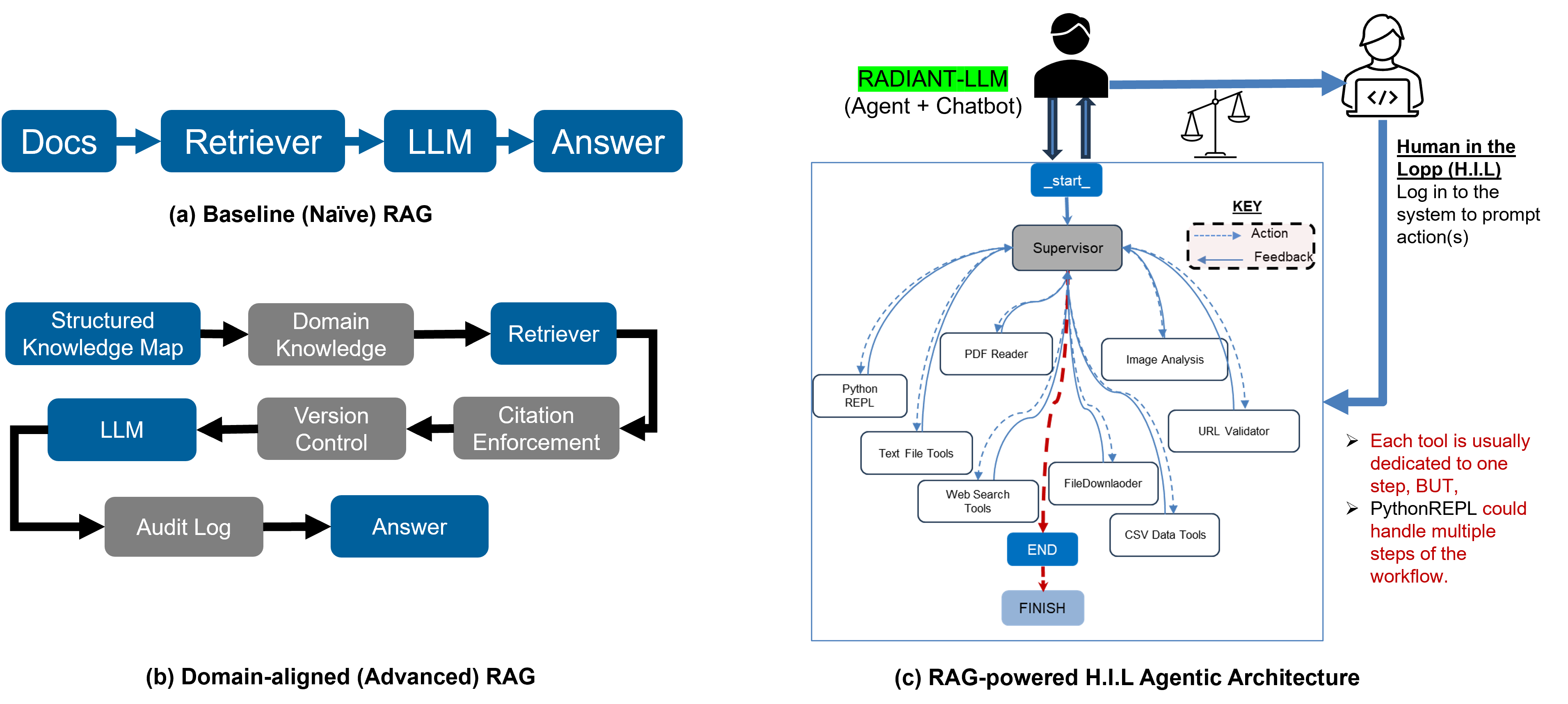}
    \caption{Architectural comparison of three RAG configurations. (a) Baseline RAG: similarity search over a flat, text only vector store. (b) Advanced RAG (this work): metadata aware retrieval using a structured knowledge map and version control to prefer authoritative sources. (c) Agentic Visual-RAG with a human in the loop (this work).}
    \label{fig:rag_architecture}
\end{figure}

Figure~\ref{fig:rag_architecture} summarizes the RADIANT-LLM retrieval pipeline. In the baseline RAG configuration (Figure~\ref{fig:rag_architecture}a), the user query is embedded and matched against a flat vector index built from text chunks. This reference pipeline does not represent document structure, modality, or metadata, so retrieved context can be weakly related to the question. In technical settings, this often manifests as incomplete answers or fabricated details, especially when the question depends on a specific figure or numerical value.

Our advanced configuration (Figure~\ref{fig:rag_architecture}b) adds a structured knowledge layer. Chunks are indexed with metadata (e.g., authors, publication date, document title, and system type), and a version gate restricts ingestion to approved, current documents. These constraints narrow retrieval to authoritative sources before generation and support the citation behavior described in Section~\ref{sec:introduction}.

The full agentic Visual-RAG configuration (Figure~\ref{fig:rag_architecture}c) replaces a fixed retrieve then generate path with a supervisor agent that selects tools based on user intent. Depending on the query, the supervisor can retrieve multimodal page level content from the PDF knowledge base, where extracted text and figure descriptions are embedded together, or query a dedicated image knowledge base that stores standalone schematics such as P\&IDs and flow diagrams. The supervisor can also inspect local CSV or Excel data, consult curated public sources, or invoke lightweight numerical analysis when computation is required.

\noindent\textbf{Human in the loop supervision:}
A human in the loop (HIL) is integrated into the execution cycle. Expert users can inspect intermediate tool outputs, verify retrieved evidence, provide corrective feedback, or override automated decisions when ambiguity is detected. When retrieval confidence is low or evidence conflicts, the supervisor surfaces candidate context and a draft rationale for review rather than proceeding automatically. Feedback is logged and can be reused in later prompts or knowledge base updates, supporting iterative refinement of retrieval and reasoning.

The retrieved multimodal context is then consolidated and passed to the LLM with system constraints that enforce citation, provenance, and refusal policies. This design supports quantitative and figure dependent queries by requiring explicit visual or structured evidence when available. When supporting evidence is missing, the system refuses to invent values, which is essential for high stakes scientific and engineering workflows.

\subsection{System Architecture: PDF Based Visual-RAG Pipeline}
\label{subsec:visual_rag_pipeline}

\begin{figure}[htbp]
\centering
\includegraphics[width=1\textwidth]{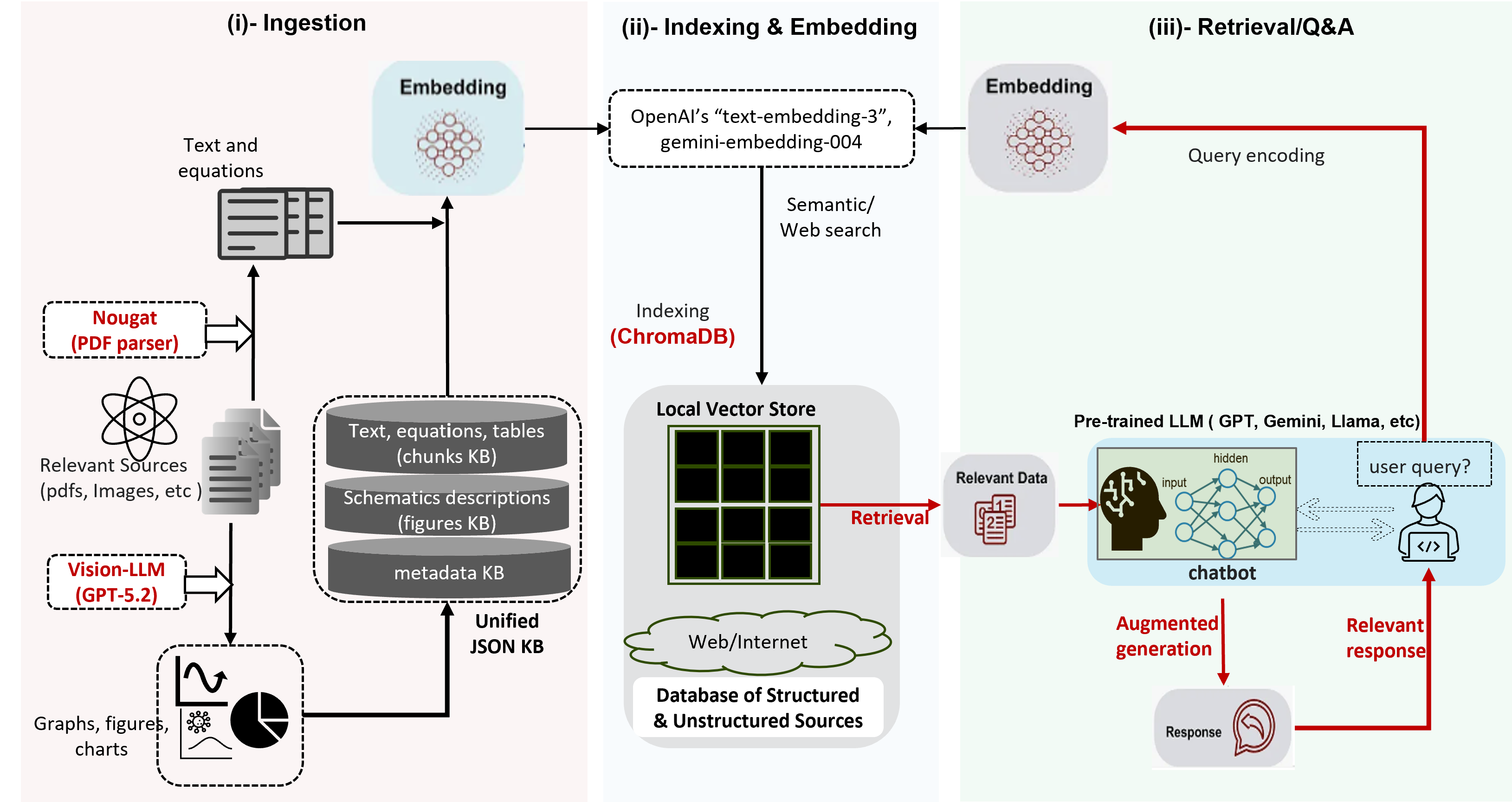}
\caption{End to end Visual-RAG pipeline in RADIANT-LLM. Documents are parsed into linked text, visual, and metadata knowledge bases (ingestion), embedded into a local vector index (embedding), and retrieved to build context for generation by a pretrained LLM (retrieval).}
\label{fig:visual_rag_architecture}
\end{figure}

RADIANT-LLM builds its knowledge base (KB) using a PDF based Visual-RAG pipeline. As shown in Figure~\ref{fig:visual_rag_architecture}, the pipeline has three stages: (i) \textit{multimodal ingestion} (red), (ii) \textit{embedding and indexing} (blue), and (iii) \textit{retrieval conditioned generation} (green). In Stage~1, each source document is parsed into text (including equations and tables), page level figure descriptions, and document metadata. In Stage~2, these outputs are stored in linked JSON KBs and embedded into a local vector index with modality tags. At query time (Stage~3), the supervisor retrieves relevant text and figure evidence and assembles the augmented prompt for the LLM. Because answer quality depends on extraction, embedding, and retrieval together, we evaluate the end to end pipeline using the metrics in Section~\ref{subsec:rag_metrics} and report results in Sections~\ref{subsec:page_level_sensitivity} and~\ref{subsec:Context_Expansion}.

We now formalize the core stages of this pipeline and highlight where RADIANT-LLM extends standard text only RAG.

\paragraph{Stage 1: Multimodal Parsing and Ingestion:}
Let $\mathcal{D} = \{D_1, D_2, \dots, D_N\}$ denote the collection of documents located in the user's secure local directory. For each document $D_i \in \mathcal{D}$, we use a hybrid parsing strategy to address the ``visual gap'' in scientific PDF processing. First, textual content (including complex equations and tables) is extracted using the Nougat library~\cite{Blecher2023Nougat:Documents}, which renders the document into a structured LaTeX or Markdown representation,
\begin{equation}
    T_i^{\text{text}} = \text{Nougat}(D_i).
    \label{eq:nougat_text_extraction}
\end{equation}
Nougat provides high fidelity OCR for text, equations, and tables, but it treats figures, plots, and schematics as whitespace. To capture the missing information, we apply a complementary visual extraction step. For each detected figure region $r \in \mathcal{R}(D_i)$, where $\mathcal{R}(D_i)$ denotes the set of figure regions in $D_i$, a vision language model (VLM) generates a structured description $c_r$ that includes quantitative and qualitative details:
\begin{equation}
    T_i^{\text{fig}} = \{\, c_r = \text{VLM}(r) \mid r \in \mathcal{R}(D_i) \,\}.
    \label{eq:vlm_figure_descriptions}
\end{equation}
In parallel, the VLM extracts document metadata such as title, authors, DOI, publication date, and source identifiers,
\begin{equation}
    M_i = \text{VLM}(D_i).
    \label{eq:metadata_extraction}
\end{equation}
The three modalities, textual content $T_i^{\text{text}}$, visual descriptions $T_i^{\text{fig}}$, and metadata $M_i$, are stored as separate but linked objects using shared document, page, and figure identifiers. This preserves modality specific detail while enabling joint retrieval across modalities, as represented in Figure~\ref{fig:visual_rag_architecture}.

\paragraph{Stage 2: Semantic Chunking and Vector Embedding:}
We apply semantic chunking only to the textual modality $T_i^{\text{text}}$ to control retrieval granularity. The text is segmented into a sequence of semantically coherent chunks $\{c_{i,1}^{\text{text}}, \dots, c_{i,N_i}^{\text{text}}\}$ with controlled length and overlap, while visual descriptions $c_r \in T_i^{\text{fig}}$ are treated as atomic units. Each textual and visual unit inherits metadata $M_i$ from its parent document, enabling citation and provenance tracking during retrieval and answer generation as illustrated in Figure~\ref{fig:rag_architecture}~(b). 

Each semantic unit is then mapped into a high-dimensional embedding space using a pretrained embedding model $E_\phi$. Textual chunks and visual descriptions are embedded separately but indexed together:
\begin{equation}
    z_{i,j}^{\text{text}} = E_\phi\!\left(c_{i,j}^{\text{text}}\right) \in \mathbb{R}^p,
    \quad
    z_{i,k}^{\text{fig}}  = E_\phi\!\left(c_{i,k}^{\text{fig}}\right) \in \mathbb{R}^p,
    \label{eq:embedding_defs}
\end{equation}
where $p$ denotes the embedding dimension. All embeddings are stored in a unified local vector index with modality tags indicating whether a vector corresponds to textual or visual content.

\paragraph{Stage 3: Multimodal Similarity Retrieval:}
Given a user query $q$, its embedding $z_q = E_\phi(q)$ is computed and compared against all stored embeddings using a standard similarity function (e.g., cosine similarity). 

\begin{equation}
    S(z_q, z_{i,m}) = \frac{z_q \cdot z_{i,m}}{\|z_q\| \, \|z_{i,m}\|} = \frac{\sum_{n=1}^{p} z_{q,n} z_{i,m,n}}{\sqrt{\sum_{n=1}^{p} z_{q,n}^2} \sqrt{\sum_{n=1}^{p} z_{i,m,n}^2}}.
\end{equation}
where $m \in \{j,k\}$.

The retriever returns a multimodal context set,
\begin{equation}
    \mathcal{C}(q) = \mathcal{C}^{\text{text}}(q) \cup \mathcal{C}^{\text{fig}}(q),
    \label{eq:multi-modal_context_set}
\end{equation}
where $\mathcal{C}^{\text{text}}(q)$ and $\mathcal{C}^{\text{fig}}(q)$ denote the top-$k$ most similar textual chunks and visual descriptions, respectively. This formulation enables direct access to quantitative and schematic information encoded in figures, plots, and diagrams, information that is typically inaccessible to text only RAG systems.

\paragraph{Stage 4: Context Aware Generation with Agentic Control:}
The retrieved context $\mathcal{C}(q)$ is combined with the user query and a system prompt $P_{\text{sys}}$, which encodes domain knowledge and response constraints. The language model then generates a response $y$ by maximizing
\begin{equation}
    P(y \mid q, \mathcal{C}(q)) = \prod_{t=1}^{T} 
    P\!\left(y_t \mid y_{<t}, q, P_{\text{sys}}, \mathcal{C}(q)\right).
    \label{eq:rag_generation_objective}
\end{equation}
A central agentic supervisor governs this process by inferring user intent from the query and orchestrating retrieval and analysis tools accordingly, as shown in Figure~\ref{fig:rag_architecture}~(c). For queries that require visual evidence, the supervisor enforces a refusal policy when
\begin{equation}
    \mathcal{C}^{\text{fig}}(q) = \emptyset,
    \label{eq:visual_refusal_condition}
\end{equation}
thereby preventing unsupported numerical or geometric claims. Through this mechanism, the system ensures that generated responses are traceable to authoritative sources and suitable for safety critical nuclear science and engineering workflows.

Equation~\eqref{eq:rag_generation_objective} retains the same autoregressive structure used by transformer LLMs~\cite{vaswani2017attention}, but conditioning on retrieved context $\mathcal{C}(q)$ and system constraints $P_{\text{sys}}$ changes what the model can plausibly generate. In standard generation, token probabilities depend only on the model's internal parameters and training data, which increases hallucination risk when relevant information is missing or underrepresented~\cite{cho2025transformer}. In the RADIANT-LLM framework, RAG anchors generation to external evidence, and the supervisor enforces availability conditions such as Equation~\eqref{eq:visual_refusal_condition}. When required evidence is absent, especially for quantitative or visual claims, the model abstains. In practice, this reduces hallucinations by constraining generation through retrieval and explicit refusal rules rather than by altering the transformer decoding algorithm.

\subsection{Quantitative Evaluation}
\label{subsec:rag_metrics}

Existing RAG benchmarks often emphasize retrieval quality or surface text similarity, but they rarely test whether an answer is correct, traceable to the right source location, and grounded in figures when needed~\cite{gao2023retrieval}. Widely used automatic metrics such as BLEU~\cite{papineni2002bleu} and ROUGE~\cite{chin2004rouge}, and question answering metrics such as Exact Match (EM) and token level F1 used in SQuAD~\cite{rajpurkar2016squad}, are useful for general language tasks. They do not directly evaluate whether a response cites the correct document and page, preserves engineering quantities, or recovers information that appears only in plots, tables, or schematics. Similar concerns have motivated domain specific GenAI evaluation frameworks in other high stakes settings such as healthcare~\cite{yi2026gai}.

In nuclear science and engineering, especially in 3S workflows, traceability, numeric fidelity, and visual fact recovery are essential. We therefore introduce a set of physics informed metrics that capture answer correctness, citation behavior, hallucination, and visual grounding in the RADIANT-LLM setting. We define an evaluation dataset consisting of $N$ query, answer, and evidence tuples:

\begin{equation}
\mathcal{Q} = \{(q_i, a_i^*, E_i^*, F_i^*)\}_{i=1}^{N},
\label{eq:eval_dataset}
\end{equation}
where $q_i$ is a user query, $a_i^*$ is the expert validated reference answer, $E_i^*$ is the minimal canonical set of authoritative evidence sources (documents, pages, or figures) used to construct and validate the query, and $F_i^*$ denotes the minimal set of visual facts that must be recovered from a schematic for visually grounded questions (empty for text only queries). Note that $E_i^*$ is not assumed to exhaust all valid supporting sources in the corpus; instead, it provides anchors for evaluation and manual verification. 

For each query, the RAG system produces an answer $a_i$ along with a set of cited sources $C_i$. The following metrics are then computed per query ($i$) and averaged over $\mathcal{Q}$. In all metric definitions, the notation $\lvert \cdot \rvert$ denotes the cardinality (i.e., the number of elements) in the set.

\vspace{0.5em}
\noindent\textbf{1. Context Precision (CoP):} 
CoP measures the correctness of the generated answer by combining semantic and numeric components. Semantic correctness is assigned discretely as
\begin{equation}
\text{CoP}_{S,i} \in \{0,\:0.25,\;0.5,\:0.75,\;1\},
\label{eq:cp_semantic}
\end{equation}
corresponding to \textit{incorrect, minimally correct, partially correct, mostly correct}, and \textit{fully correct} answers, respectively, based on expert judgment. For answers containing numeric facts, such as labeled neutron fluxes, heat rates, or distances, numeric correctness is computed as
\begin{equation}
\text{CoP}_{N,i} =
\frac{1}{m_i}
\sum_{j=1}^{m_i}
\left(
1 - \frac{|v_{i,j} - v^*_{i,j}|}{\max(|v^*_{i,j}|,\epsilon)}
\right),
\label{eq:cp_numeric}
\end{equation}
where $v_{i,j}$ is a numeric value extracted from the generated answer, $v^*_{i,j}$ is the corresponding reference value, $m_i$ is the number of numeric facts, and $\epsilon$ is a small constant preventing division by zero. When multiple numeric values are generated for a given reference quantity, we select the value that maximizes the numeric fidelity term in Equation~\eqref{eq:cp_numeric}; if no corresponding value is present, that term contributes zero.

The final CoP score is defined in a piecewise manner to account for the presence or absence of semantic and numeric components:

\begin{equation}
\text{CoP}_i =
\begin{cases}
\text{CoP}_{N,i}, & \text{if } \text{CoP}_{S,i} \text{ is undefined}, \\[0.5em]
\text{CoP}_{S,i}, & \text{if } \text{CoP}_{N,i} \text{ is undefined}, \\[0.5em]
\alpha\,\text{CoP}_{S,i} + (1-\alpha)\,\text{CoP}_{N,i}, & \text{otherwise},
\end{cases}
\label{eq:cp_def}
\end{equation}
where $\alpha \in [0,1]$ controls the relative weight of semantic versus numeric fidelity. Lower values of $\alpha$ place more emphasis on numeric accuracy (as might be preferred in detailed reactor physics and facility design calculations), whereas higher values favor semantic correctness and explanation quality (as may be appropriate in safeguards or policy oriented assessments). In this work, $\alpha$ is tuned based on the nature of the question; for most mixed semantic and numeric queries, we use $\alpha = 0.6$, which slightly favors semantic correctness while still giving substantial weight to numerical accuracy.

\vspace{0.5em}
\noindent\textbf{2. Citation Precision (CiP):} 
CiP evaluates whether the sources cited by the system support the generated answer. For each query $i$, the model produces a set of citations $C_i$. Each citation specifies either a link (online sources) or a document ID for local sources and, when available, a page, section, or figure. Human evaluators inspect the cited locations and label each citation as either \emph{valid} (the cited location substantively supports at least one claim in the answer $a_i$) or \emph{invalid} (no such supporting information is present). 

For a given query $i$,
$
\text{CiP}_i = \frac{\#\text{ valid citations in } C_i}{\max(\#\text{ citations generated}, 1)}.$
Formally,
\begin{equation}
\text{CiP}_i =
\frac{
\left|
\left\{ c \in C_i \;:\; c \text{ is judged valid by a human evaluator} \right\}
\right|
}
{\max\left(|C_i|, 1\right)}.
\label{eq:cip_def}
\end{equation}
Here $\text{CiP}_i = 1$ indicates that all citations produced by the model are judged correct and supporting. When $C_i$ is empty, $\text{CiP}_i$ is defined as $0$ unless no citations are required for that query (e.g., purely definitional questions), in which case $\text{CiP}_i$ is ignored in the average statistics.

\vspace{0.5em}
\noindent\textbf{3. Citation Hit (CiH):}
CiH measures whether the model successfully anchors its answer to at least one canonical evidence item in $E_i^*$ for a given query. Unlike CiP, which evaluates the correctness of \emph{all} generated citations, CiH is a coarse, binary indicator of whether the answer is grounded in at least one of the expert-identified sources used to construct the question. For queries where $E_i^* \neq \varnothing$, CiH is defined as
\begin{equation}
\text{CiH}_i =
\begin{cases}
1, & \text{if } \exists\, c \in C_i,\, e \in E_i^* \text{ such that } c \text{ is judged to refer to } e, \\[0.5em]
0, & \text{otherwise},
\end{cases}
\label{eq:cih_def}
\end{equation}
where we say that a citation $c$ refers to $e$ when, under human inspection, $c$ explicitly points to the same document (and, when specified, the same page, section, or figure) as the canonical anchor $e$. For queries where $E_i^* = \varnothing$ (no canonical evidence anchors are required), CiH is not computed and is excluded from average statistics.

\vspace{0.5em}
\noindent\textbf{4. Hallucination Rate (HR):}
HR measures the extent to which a model introduces unsupported content in its generated answer. For a given query $i$, the model output $a_i$ is decomposed into a set of \emph{atomic generated claims}, denoted $\mathcal{K}_{\text{gen}}(a_i)$. Let $\mathcal{K}_{\text{unsupported}}(a_i, C_i)$ denote the subset of generated claims not supported by the available context. The Hallucination Rate is then defined as
\begin{equation}
\text{HR}_i =
\frac{\left|\mathcal{K}_{\text{unsupported}}(a_i, C_i)\right|}
{\left|\mathcal{K}_{\text{gen}}(a_i)\right|}.
\label{eq:hr_def}
\end{equation}

In practice, LLM responses may contain multiple claims, some of which may not be strictly required to answer the query. In this work, a claim is considered \emph{unsupported} if it is factually incorrect (reduces CoP) or cannot be directly retrieved or reasonably inferred from the available visual or textual evidence, independent of whether an explicit citation marker is present. Extra geometric, physical, or mathematical inferences that are consistent with the given context are not penalized. Lower HR indicates a more faithful answer. HR captures unsupported content generation and is evaluated independently of citation fidelity, which is assessed separately via CiP and CiH.

\vspace{0.5em}
\noindent\textbf{5. Visual Recall (ViR):} 
For visually grounded queries, ViR evaluates the system's ability to recover the \emph{minimal set of visual facts} from figures or schematics required to answer the query. Let $F_i$ be the set of visual facts contained in the generated answer. ViR is defined as
\begin{equation}
\text{ViR}_i =
\frac{|F_i \cap F_i^*|}{|F_i^*|}.
\label{eq:vra_def}
\end{equation}
Fact matching is exact for quantitative visual facts (e.g., distances, radii, labeled values), as well as for symbolic, categorical, and relational visual labels. ViR measures recall over the required visual facts expressed in the \emph{generated answer}. This is independent of downstream reasoning or numeric derivation, and is computed only for queries that explicitly require figure or schematic based evidence.

Note that ViR is inherently bound to the quality of the VLM used during KB construction (see Equation~\eqref{eq:vlm_figure_descriptions}). Lower capability VLMs may fail to extract all numerical annotations from complex technical schematics, resulting in a reduced ViR score despite correct retrieval and generation. This behavior is shown in the results (Section~\ref{subsec:page_level_sensitivity}).

\vspace{0.5em}
\noindent\textbf{Average Metrics:} 
System level performance is reported as the mean value of each metric over the entire set of $N$ query/answer/evidence tuples:
\vspace{0.5em}

\begingroup
\scriptsize
\begin{equation}
\overline{\text{CoP}} = \frac{1}{N}\sum_{i=1}^{N} \text{CoP}_i,
\quad
\overline{\text{CiP}} = \frac{1}{N}\sum_{i=1}^{N} \text{CiP}_i,
\quad
\overline{\text{CiH}} = \frac{1}{N}\sum_{i=1}^{N} \text{CiH}_i,
\quad
\overline{\text{HR}} = \frac{1}{N}\sum_{i=1}^{N} \text{HR}_i,
\label{eq:average_metrics}
\end{equation}
\endgroup
with $\overline{\text{ViR}}$ computed over the subset of visual queries only. Note that since CiH is binary at the query level, its mean value represents the fraction of queries for which the model successfully anchored its response to at least one expert validated evidence source. 

Together, these metrics provide a compact framework for evaluating multimodal, citation focused Visual-RAG relative to text only RAG or generic document chat models. Instead of collapsing evaluation into a single score, we separate (i) semantic and numeric correctness (CoP), (ii) citation precision and anchor hits (CiP, CiH), (iii) unsupported content (HR), and (iv) recovery of required visual facts (ViR).

\vspace{0.5em}
\noindent\textbf{Expert verification vs. automatic string matching.}
During metric design, we tested automatic schemes based on citation strings and embedding similarity (e.g., cosine similarity between citation anchors, fuzzy matching of page and section labels, and Analytic Hierarchy Process (AHP) weighting to combine metrics into a single composite score). In practice, these approaches were brittle with respect to formatting variations, paraphrasing, and nearby but incorrect references. For a single canonical anchor (e.g., \texttt{1167010.pdf, p.~6}) and several nearby incorrect locations (e.g., \texttt{1167010.pdf, p.~13, Sec.~3.1}, \texttt{p.~13, Sec.~3.2}, \texttt{p.~9 to 10}), similarity scores were typically $0.64$ to $0.71$. No fixed threshold separated correct from incorrect citations while materially reducing expert effort. For these reasons, we discarded fully automatic formulations in favor of simpler, expert verified definitions for CoP, CiP, CiH, HR, and ViR.



\subsection{Auxiliary Tools for the Visual-RAG Pipeline}
\label{subsec:multi-modal_aux_pipelines}

The PDF based Visual-RAG pipeline is the primary knowledge ingestion and retrieval mechanism in RADIANT-LLM, but the same agentic retrieval principles extend to additional data modalities through the auxiliary tools summarized in Table~\ref{tab:radiant_tools}. The supervisor agent selects these tools at runtime to retrieve, combine, and validate information from multiple knowledge sources based on the query intent and requirements.

\begin{table}[H]
\centering
\small
\caption{Custom tool suite used by RADIANT-LLM. The supervisor agent selects these tools to execute workflows while keeping raw data local.}
\label{tab:radiant_tools}
\resizebox{\textwidth}{!}{%
\begin{tabular}{|l|p{10cm}|}
\hline
\textbf{Tool Name} & \textbf{Description} \\
\hline
\texttt{PDFReaderTool} (Core) & Main RAG engine for documents. Parses PDFs using text and vision models, updates the local JSON KB with text, equations, tables, figure descriptions, and metadata, embeds content, and performs semantic search. \\
\hline
\texttt{ImageAnalysisTool} & Vision focused pipeline that extracts technical information from standalone images (plots, schematics, flow diagrams) and stores textual descriptions and tags in the JSON KB. \\
\hline
\texttt{CSVDataQueryTool} & Performs focused RAG style queries on CSV or Excel files, complementing the capabilities of \texttt{CSVDataFinderTool}. \\
\hline
\texttt{Web Search Tools} & Optional tools that connect to public search engines and Wikipedia for current retrieval of non sensitive information and references. Results can be inspected and, if appropriate, added to the local KB. \\
\hline
\texttt{FileDownloaderTool} & Retrieves documents from open repositories (e.g., OSTI or NRC ADAMS) into the secure local directory, where they are ingested by the PDF pipeline. \\
\hline
\texttt{PythonREPLTool} & Executes Python code for calculations, simple data analysis, and scripted operations on files within the working directory. \\
\hline
\texttt{LightWeightPDFReaderTool} & A faster but less detailed PDF processor that can be used when full multimodal parsing is not required. \\
\hline
\texttt{LightWeightCSVDataTool} & Loads structured data from CSV/Excel files, allowing the agent to answer quantitative and aggregation queries over tabular datasets. \\
\hline
\end{tabular}%
}
\end{table}


Standalone images (e.g., plant layouts, piping and instrumentation diagrams (P\&IDs), and flow diagrams) are handled by the \texttt{ImageAnalysisTool}, which mirrors the PDF workflow. Each image is processed by a VLM (Equation~\eqref{eq:vlm_figure_descriptions}) to produce a single structured description capturing geometry, labels, and quantitative annotations. The description is treated as an \emph{atomic semantic unit} linked to the image file, enriched with metadata (file name, inferred topic, component type, and optional user annotations), and embedded into a dedicated image knowledge base. Because this description already represents the smallest coherent unit of visual meaning, no further chunking is applied. At query time, these image embeddings can be retrieved alongside PDF derived content within the same similarity search framework.

Structured numerical data stored in CSV or Excel files is handled by the \texttt{CSVDataQueryTool} and \texttt{LightWeightCSVDataTool}. Rather than embedding entire tables into the vector store, these tools load datasets into data frames and interpret user queries as computational operations over the data (e.g., filtering, aggregation, or simple statistical analysis). The supervisor typically generates short Python code to execute these operations, and the resulting values can be combined with document or image evidence during answer synthesis.

For web queries, the supervisor may invoke web search tools to retrieve current, non sensitive information. In contrast to local PDF and image pipelines, web content is not persistently ingested or embedded by default. Instead, a small number of relevant snippets is extracted and passed directly to the LLM for immediate contextual grounding. This design preserves a clear boundary between user approved authoritative sources and opportunistic external data.

Across modalities, the supervisor can retrieve from one or more knowledge bases as required by the query. For example, a question concerning decay heat behavior for a specific fuel type may trigger retrieval from the PDF KB (standards text), the image KB (decay heat plots), and a CSV dataset (numerical values), with the agent integrating these evidence streams into a single, well cited response. In this way, the auxiliary tools extend the Visual-RAG formulation beyond documents while keeping data local and maintaining the same citation and provenance requirements.

\subsection{Data Sourcing and Curation}
\label{subsec:data_sourcing}

For this study, open access documents from \textit{OSTI.gov} were used as the main public corpus. A CSV file containing metadata for 1{,}000 candidate documents (titles, authors, dates, OSTI IDs, descriptions, URLs, and related fields) was used to guide selection and ingestion. Although the \texttt{FileDownloaderTool} is intended for general purpose downloading from public repositories, a dedicated utility was implemented to interface with the OSTI.gov API and retrieve the selected documents in bulk, reducing manual effort and avoiding unnecessary LLM calls during data collection. Of the 1{,}000 OSTI entries, 984 files were successfully retrieved; 16 links were either corrupted or returned ``Not Found.'' Among the retrieved files, 271 were non PDF (e.g., HTML pages) and were excluded from the PDF based RAG pipeline.

From the remaining PDFs, multiple knowledge bases were constructed to demonstrate different use cases. First, a focused 3S KB was created by sampling 250 representative documents and processing them with the \texttt{PDFReaderTool}. This KB contained regulatory, safeguards, and security relevant material and was used to test RADIANT-LLM on safety, security, and safeguards questions. Second, the remaining OSTI PDFs were embedded into a broader nuclear science and engineering KB aimed at more general question answering tasks. Third, additional proprietary and unpublished materials (e.g., internal reports and draft analyses) were ingested into separate local KBs to demonstrate RADIANT-LLM's ability to integrate and manage private knowledge alongside public sources. Because the ingestion and curation steps are modular, the same process can be extended to other authoritative sources such as the U.S. NRC, the IAEA, digital libraries of conference proceedings, and expert curated Q\&A repositories.

Beyond bulk ingestion, the framework also supports dynamic KB creation starting from a small set of seed documents. Under user control, the agent can extract references from existing PDFs, use web search and downloading tools to locate publicly available versions of cited reports or standards, and then call the \texttt{PDFReaderTool} to parse and embed these new documents into the chosen KB. In this way, an initial minimal corpus can be expanded iteratively into a richer, task specific KB without changing the underlying model or deployment setup.

\subsection{Data Security and Privacy}
\label{subsec:data_security}

Data security and privacy are central to the design of RADIANT-LLM, especially for proprietary designs, sensitive 3S analyses, and export controlled information. The framework is independent of the underlying model and deployment platform: it can run with different pretrained LLMs and can be installed on local workstations or institutional servers. In all cases, raw documents (PDFs, images, CSV or Excel files), parsed text, image descriptions, embeddings, and JSON knowledge bases remain on the user's infrastructure. The system operates on user specified directories and stores all intermediate and derived data locally. When an external LLM is used via API, only compact prompts are transmitted, consisting of the user's question and the minimal retrieved excerpts needed to answer it, not the full documents. The same workflow can be paired with fully local LLMs, in which case no external calls are required.

Within this setup, users retain full control over KB management. They decide which directories are exposed, which public sources are trusted, and how KBs are partitioned (e.g., separating public standards from facility specific material or unpublished work). Web search and other outward facing tools are optional and can be restricted to non sensitive queries, with retrieved snippets used transiently and excluded from the persistent KB unless explicitly approved. Dynamic KB expansion, such as adding new documents discovered through reference chasing or web search, is initiated and confirmed by the user before ingestion. Combined with the citation focused RAG pipeline described in Section~\ref{sec:methods}, these measures support safety, security, and safeguards requirements without compromising confidentiality or data sovereignty.

\subsection{Sensitivity Analysis and Benchmark Design}
\label{subsec:benchmark_design}

To evaluate the proposed Visual-RAG architecture, we adopted a task driven, expert curated sensitivity analysis rather than a large crowdsourced dataset. The overall benchmark comprised two complementary components: (i) a \emph{foundational page level visual sensitivity benchmark} on representative technical pages, and (ii) a \emph{retrieval sensitivity analysis under context scaling}, using document level and corpus level knowledge bases (KBs) constructed from 3S regulatory and technical reports. For each benchmark query, generated answers were evaluated against expert curated reference data using the tuple $\{q_i, a_i^*, E_i^*, F_i^*\}$ defined in Equation~\eqref{eq:eval_dataset} and the metrics in Section~\ref{subsec:rag_metrics}. For brevity, this paper reports only diagnostic metrics and qualitative failure modes; the verbatim questions, answer rubrics, and YAML scoring templates will be released separately on our public code repository.

For the \textit{page level benchmark}, three technical pages (calculus and reactor physics) containing both text and dense visual content (schematics, plots, and labeled dimensions) were sampled: one from a classical calculus text~\cite{MarchWolff1917_Calculus} and two from \emph{Nuclear Systems I: Thermal Hydraulic Fundamentals}~\cite{Todreas1989NuclearFundamentals}. Each page was ingested using the RADIANT-LLM parser, which combines Nougat for text, equation, and table extraction with a vision language model (VLM) for visual parsing (Figure~\ref{fig:visual_rag_architecture}). Separate KBs were constructed using different VLM configurations from the ``GPT-4.x'' and ``GPT-5.x'' families. Five expert curated questions were constructed per page (15 total), spanning figure only, text only, and cross modal queries, and evaluated using CoP, HR, and ViR. Citation metrics (CiP and CiH) were de emphasized here because each KB contained a single page, causing CiP and CiH to saturate at 1. These pages were chosen as canonical nuclear science and engineering examples rather than 3S content to decouple visual reasoning performance from regulatory context and to serve as a controlled unit test of multimodal ingestion and retrieval fidelity. Based on these page level results, GPT-5.2 was selected as the default VLM configuration for all subsequent large KB construction.

For the \textit{context expansion (scaling) benchmark}, the same evaluation tuple was applied while expanding KB scope across regimes of increasing complexity. In the \emph{document level} (single source) regime, the KB contained a single full technical report. Specifically, a design document for a used nuclear fuel storage facility (UNFSF)~\cite{badwan2015application} was chosen, and questions were sampled from specific pages within that report. In the \emph{corpus level} (multi source) regime, the target document was embedded within collections of 5, 10, 20, \ldots, up to 250 additional sources drawn from 3S related regulatory and technical reports. In these settings, the focus shifted from pure visual grounding to retrieval discrimination and citation behavior under semantic redundancy and retrieval competition. Accordingly, CiP, CiH, HR, and CoP were used as the primary indicators of reliability under context scaling, with ViR computed for the subset of visually grounded queries. All KBs in this benchmark were constructed using the selected GPT-5.2 VLM for figures and the Nougat based parser for text, equations, and tables. Results from the page level and context expansion sensitivity analyses are reported in Sections~\ref{subsec:page_level_sensitivity} and~\ref{subsec:Context_Expansion}, respectively.


\section{Results and Discussion}
\label{sec:results}


\subsection{Multimodal Ingestion and Visual Semantic Recovery}
\label{subsec:ingestion}

\begin{figure*}[htbp]
    \centering
    \includegraphics[width=\textwidth]{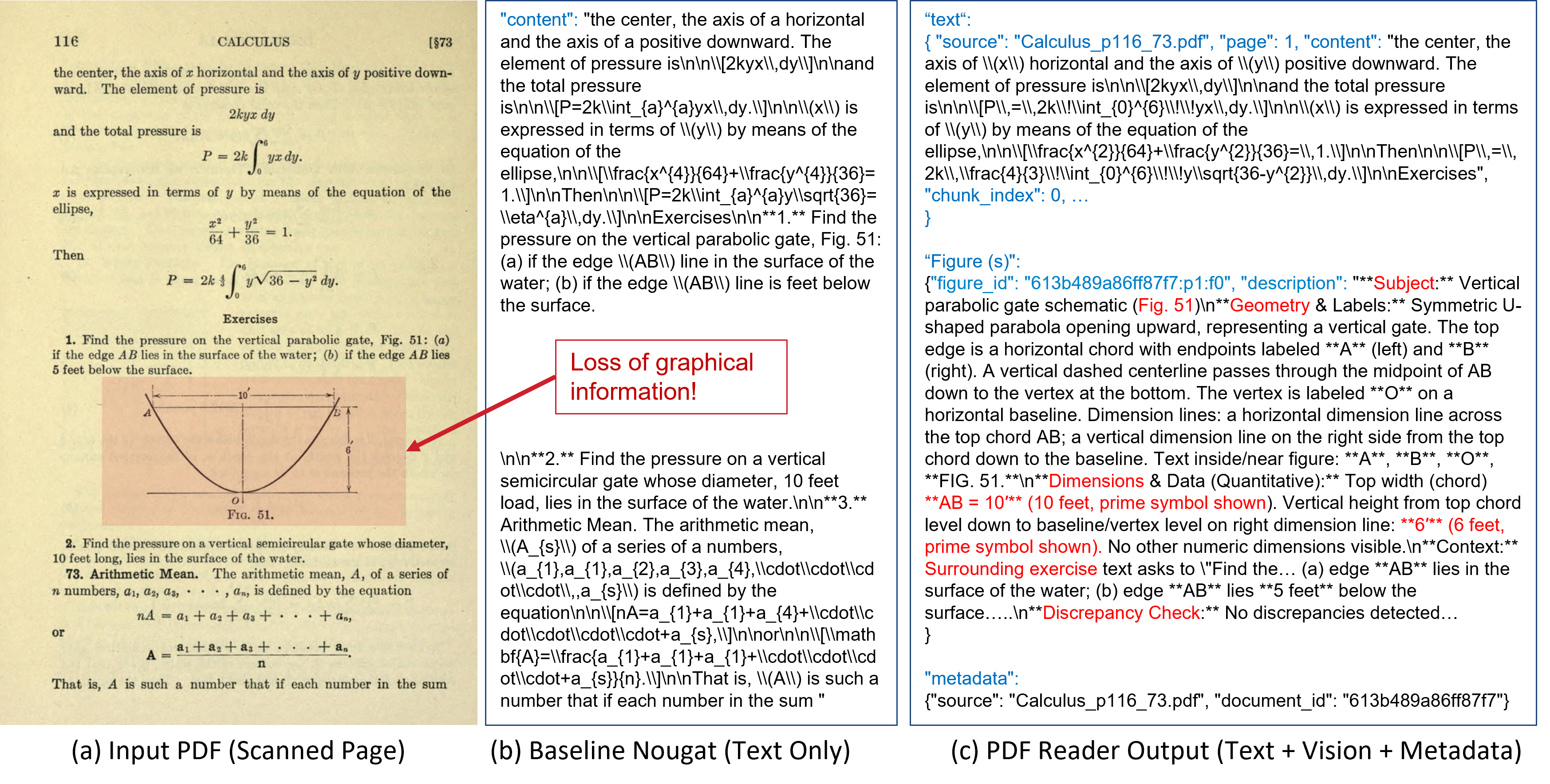}
    \caption{Comparison of text only and multimodal PDF parsing on a calculus page. (a) Original PDF page~\cite{MarchWolff1917_Calculus} showing text, equations, and a figure. (b) Output from the Nougat parser showing loss of visual information (modified from Figure~B.1 in~\cite{blecher2023nougat}). (c) RADIANT-LLM parsing output, where a vision language model recovers the diagram semantics (labels and topology) into a retrievable JSON representation.}
    \label{fig:nougat_comparison}
\end{figure*}

To illustrate visual information loss during ingestion, we repeated the extraction example from the Nougat paper using the same calculus page (Figure~B.1 in~\cite{blecher2023nougat}), which later served as Case~1 in our page level benchmark. As shown in Figure~\ref{fig:nougat_comparison}, the Nougat parser extracts surrounding text and equations but omits the embedded schematic entirely, leaving a blank region where the figure appeared in the original document. This reflects a common limitation of text centric OCR pipelines: geometry, dimensions, and boundary conditions that live in figures are not ingested into the knowledge base.

In RADIANT-LLM, ingestion applies VLM based parsing during preprocessing, converting the schematic into a structured textual description before embedding. The recovered representation captures geometric relationships and labeled dimensions present only in the figure, enabling these facts to be retrieved and cited in downstream queries. Compared with approaches that crop or splice figures without semantic interpretation, the page level multimodal ingestion preserves local context from captions and adjacent text, reducing ambiguity and limiting hallucination.

This experiment focuses on ingestion: whether the KB contains the visual facts required to answer engineering questions grounded in schematics and annotated diagrams. We next report the page level retrieval sensitivity study.

\subsection{Page Level Retrieval Sensitivity}
\label{subsec:page_level_sensitivity}

This section reports the performance of the Visual-RAG architecture on the page level sensitivity benchmark defined in Section~\ref{subsec:benchmark_design}. 

\subsubsection{Mean Page Level Metric Scores}
\label{subsubsection: mean_page_level_metrics}

\begin{figure}[htbp]
  \centering
  \includegraphics[width=\textwidth]{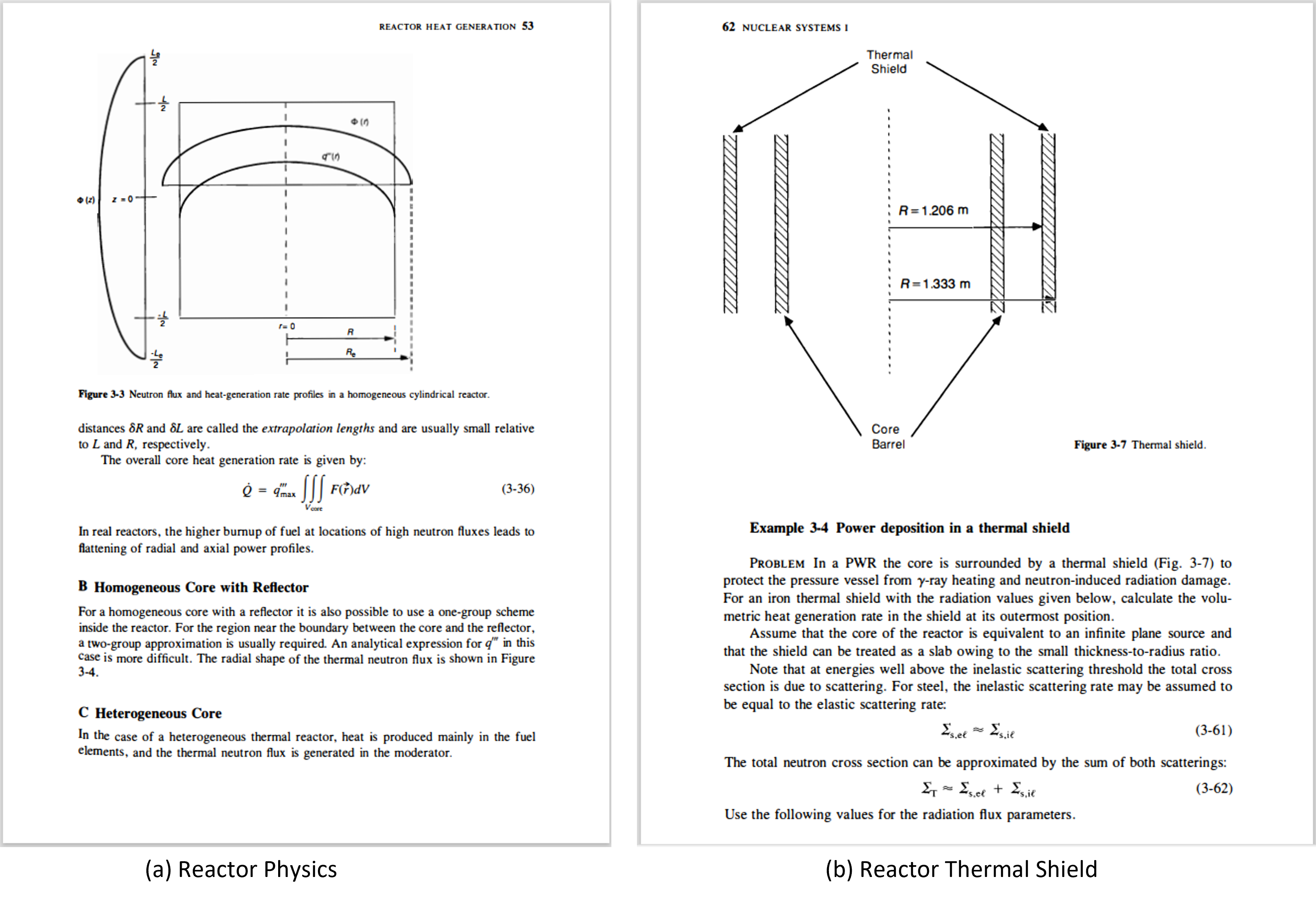}
  \caption{Technical pages used in the page level benchmark. Left: homogeneous cylindrical reactor neutron flux and heat generation profiles (Figure~3--3). Right: thermal shield geometry and power deposition model (Figure~3--7). These pages contain dense visual annotations and served as sources for figure only, text only, and hybrid queries.}
  \label{fig:page_level_sources}
\end{figure}

The calculus page (Figure~\ref{fig:nougat_comparison}) concentrates key geometric information in a single schematic, while the two reactor physics pages from \emph{Nuclear Systems I}~\cite{Todreas1989NuclearFundamentals}, shown in Figure~\ref{fig:page_level_sources}, depict neutron flux and heat generation profiles (Figure~3--3) and power deposition in a thermal shield geometry (Figure~3--7), respectively.

\begin{figure*}[htbp]
    \centering
    \begin{subfigure}[t]{0.32\textwidth}
        \centering
        \includegraphics[width=\textwidth]{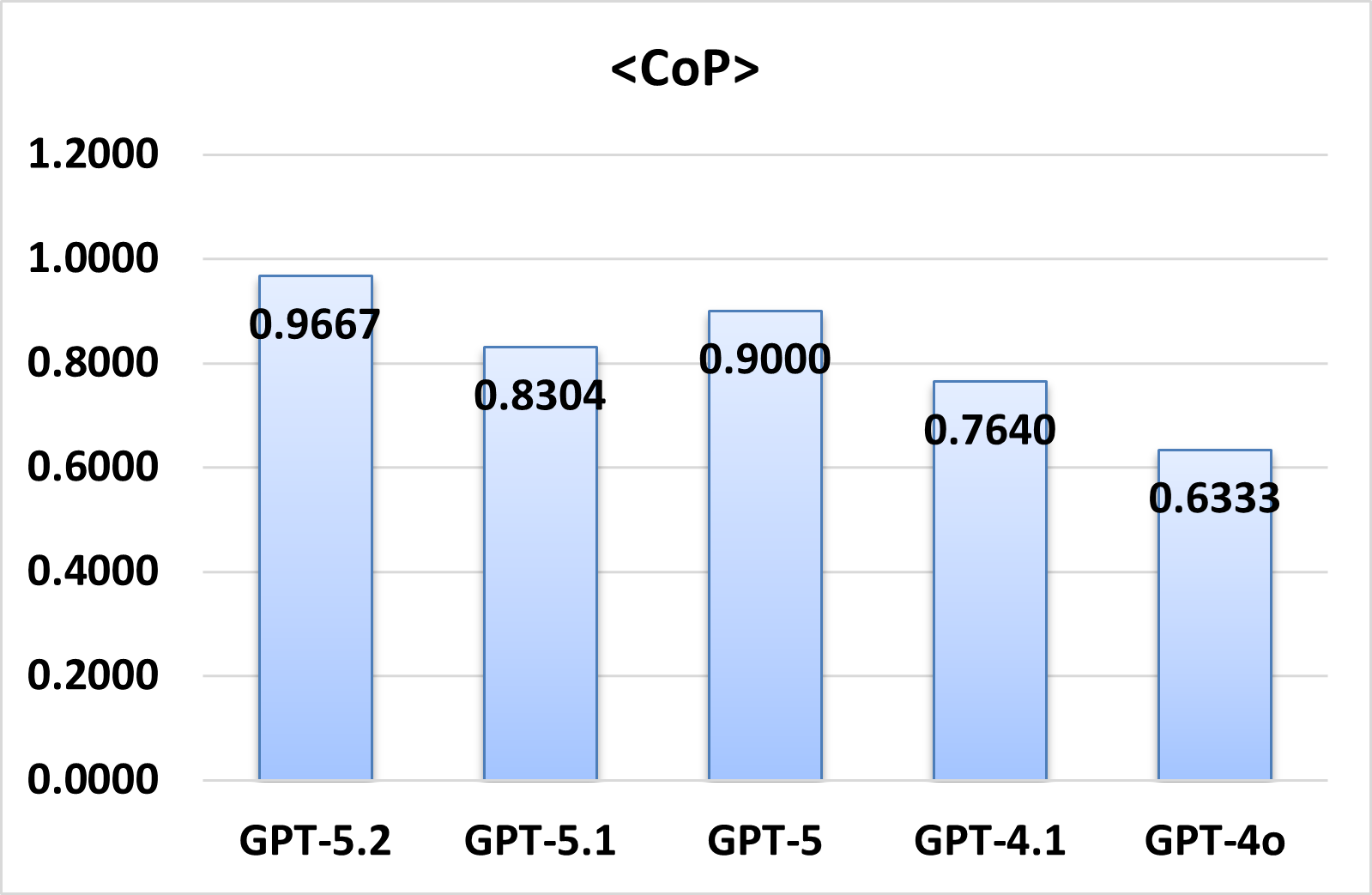}
        \caption{CoP ($\overline{\mathrm{CoP}}$)}
        \label{fig:page_cp}
    \end{subfigure}
    \hfill
    \begin{subfigure}[t]{0.32\textwidth}
        \centering
        \includegraphics[width=\textwidth]{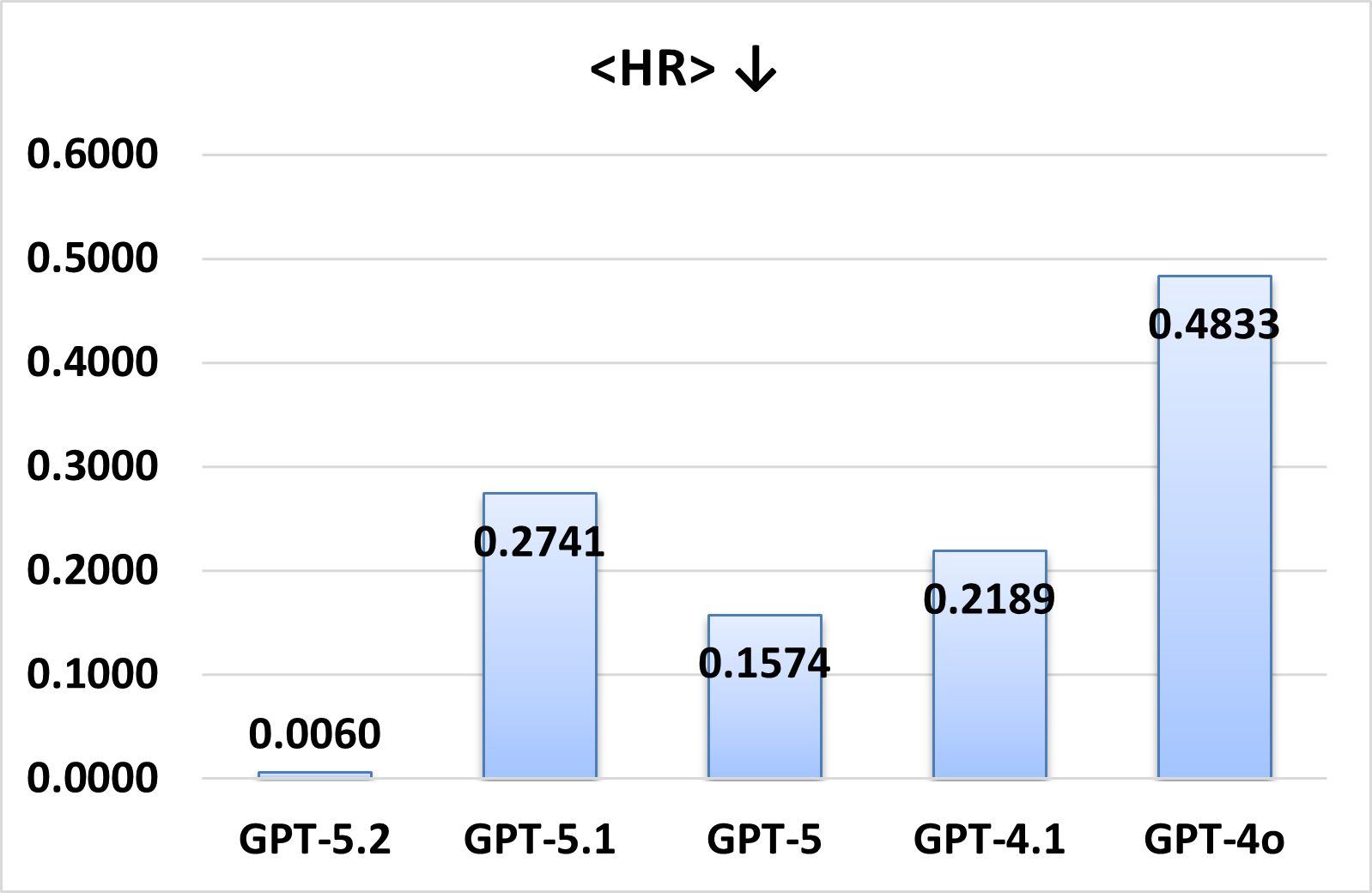}
        \caption{HR ($\overline{\mathrm{HR}}\downarrow$)}
        \label{fig:page_hr}
    \end{subfigure}
    \hfill
    \begin{subfigure}[t]{0.32\textwidth}
        \centering
        \includegraphics[width=\textwidth]{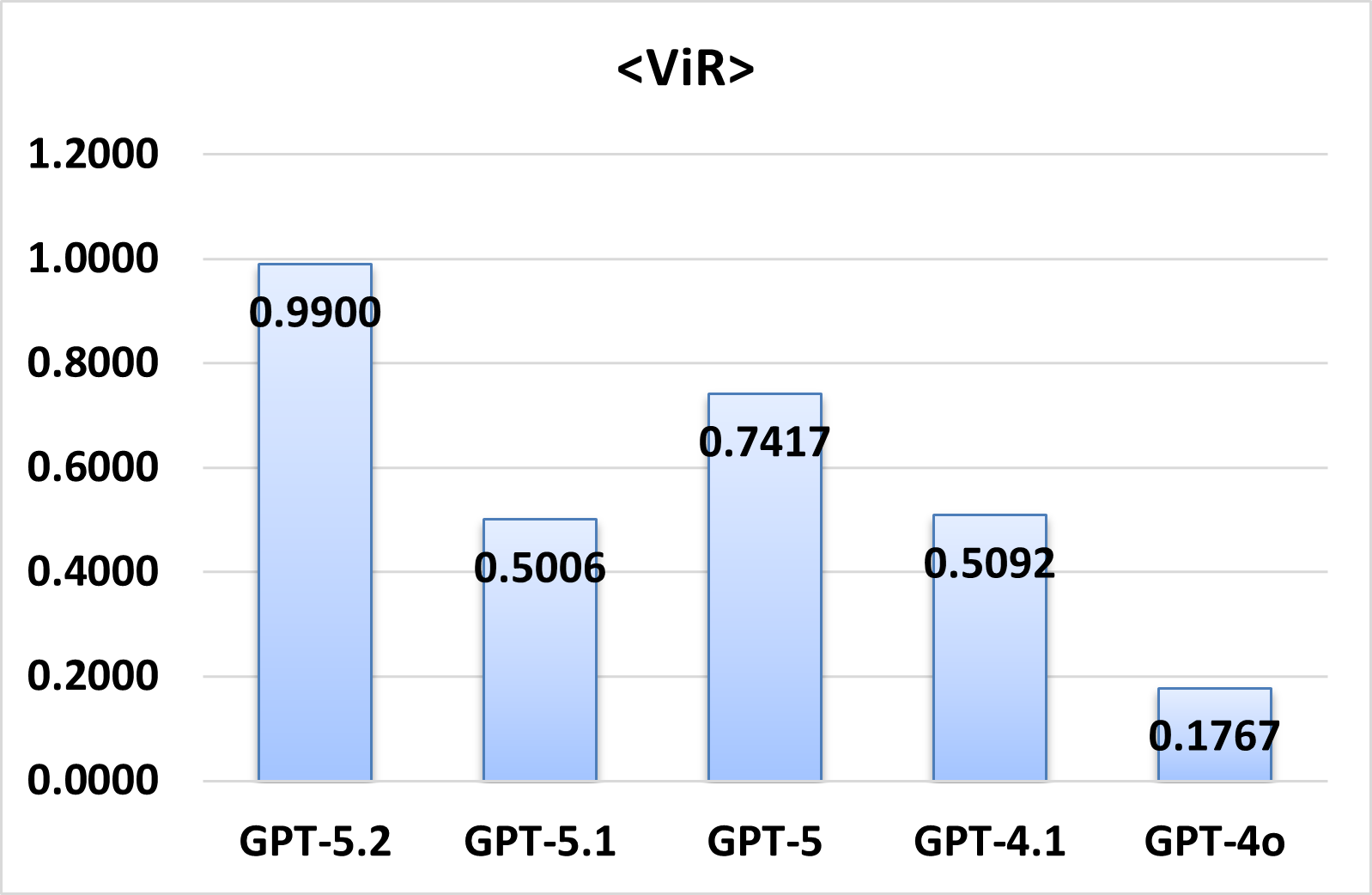}
        \caption{ViR ($\overline{\mathrm{ViR}}$)}
        \label{fig:page_vra}
    \end{subfigure}

    \caption{Page level benchmark results averaged over 15 queries. Shown are mean CoP ($\overline{\mathrm{CoP}}$), HR ($\overline{\mathrm{HR}}$, lower is better), and ViR ($\overline{\mathrm{ViR}}$) for GPT-5.2, GPT-5.1, GPT-4.1, and GPT-4o. Higher ViR generally coincides with higher CoP and lower HR, indicating the importance of visual grounding for factual correctness.}
    \label{fig:page_level_bars}
\end{figure*}

Table~\ref{tab:per_case_averages} and Figure~\ref{fig:page_level_bars} summarize performance across all 15 queries. The GPT-5.2 configuration achieved the highest scores across all three cases, with $\overline{\mathrm{CoP}}=0.9667$, $\overline{\mathrm{HR}}=0.0060$, and near perfect visual recall ($\overline{\mathrm{ViR}}=0.9900$). Using GPT-5 as the reasoning model also yielded strong performance ($\overline{\mathrm{CoP}}=0.9000$, $\overline{\mathrm{ViR}}=0.7417$) but with a higher hallucination rate ($\overline{\mathrm{HR}}=0.1574$), indicating more frequent unsupported statements. GPT-5.1 and GPT-4.1 showed intermediate behavior (moderate $\overline{\mathrm{CoP}}=0.8304$ and $0.7640$, with $\overline{\mathrm{ViR}}\approx 0.50$), whereas GPT-4o displayed the weakest visual grounding ($\overline{\mathrm{ViR}}=0.1767$) and the highest hallucination rate ($\overline{\mathrm{HR}}=0.4833$).

Across all configurations, higher $\overline{\mathrm{ViR}}$ was consistently associated with higher $\overline{\mathrm{CoP}}$ and substantially lower $\overline{\mathrm{HR}}$. This indicates that overall performance is governed less by generic text generation capability than by the VLM's ability to extract and generalize key schematic details into accurate textual representations that can be embedded and retrieved. When the VLM faithfully captures geometric relationships, profile shapes, coordinate labels, and other factual elements from the figures, downstream QA is more reliably grounded in the correct page and figure evidence, which in turn suppresses hallucinations. CiP and CiH saturated at or near unity for most models, confirming that CoP, HR, and ViR were the most discriminating metrics in this setting.

\begin{table*}[htbp]
\centering
\scriptsize
\renewcommand{\arraystretch}{1.15}
\caption{Overall average performance across page level Visual-RAG queries.
Reported metrics include CoP, HR (lower is better), and ViR. Averages are computed over three cases (15 queries total).}
\label{tab:per_case_averages}
\begin{tabular}{llccccc}
\toprule
\textbf{Model} & \textbf{Case} & \textbf{CoP} & \textbf{CiP} & \textbf{CiH} & \textbf{HR$\downarrow$} & \textbf{ViR} \\
\midrule
GPT-5.2 & Calculus (Q1--5) & 1.0000 & 1.0000 & 1.0000 & 0.0000 & 1.0000 \\
GPT-5.2 & Reactor Physics (Q6--10) & 0.9000 & 1.0000 & 1.0000 & 0.0180 & 0.9700 \\
GPT-5.2 & Thermal Shield (Q11--15) & 1.0000 & 1.0000 & 1.0000 & 0.0000 & 1.0000 \\
GPT-5.2 & \textbf{Average (15Q)} & \textbf{0.9667} & \textbf{1.0000} & \textbf{1.0000} & \textbf{0.0060} & \textbf{0.9900} \\
\midrule
GPT-5.1 & Calculus (Q1--5) & 0.7911 & 1.0000 & 1.0000 & 0.3500 & 0.2767 \\
GPT-5.1 & Reactor Physics (Q6--10) & 0.7000 & 1.0000 & 1.0000 & 0.4056 & 0.2917 \\
GPT-5.1 & Thermal Shield (Q11--15) & 1.0000 & 1.0000 & 1.0000 & 0.0667 & 0.9333 \\
GPT-5.1 & \textbf{Average (15Q)} & \textbf{0.8304} & \textbf{1.0000} & \textbf{1.0000} & \textbf{0.2741} & \textbf{0.5006} \\
\midrule
GPT-5 & Calculus (Q1--5) & 1.0000 & 1.0000 & 1.0000 & 0.0000 & 1.0000 \\
GPT-5 & Reactor Physics (Q6--10) & 0.7000 & 1.0000 & 1.0000 & 0.4056 & 0.2917 \\
GPT-5 & Thermal Shield (Q11--15) & 1.0000 & 1.0000 & 1.0000 & 0.0667 & 0.9333 \\
GPT-5 & \textbf{Average (15Q)} & \textbf{0.9000} & \textbf{1.0000} & \textbf{1.0000} & \textbf{0.1574} & \textbf{0.7417} \\
\midrule
GPT-4.1 & Calculus (Q1--5) & 0.5920 & 1.0000 & 1.0000 & 0.4667 & 0.0833 \\
GPT-4.1 & Reactor Physics (Q6--10) & 0.7000 & 1.0000 & 1.0000 & 0.1900 & 0.4444 \\
GPT-4.1 & Thermal Shield (Q11--15) & 1.0000 & 1.0000 & 1.0000 & 0.0000 & 1.0000 \\
GPT-4.1 & \textbf{Average (15Q)} & \textbf{0.7640} & \textbf{1.0000} & \textbf{1.0000} & \textbf{0.2189} & \textbf{0.5092} \\
\midrule
GPT-4o & Calculus (Q1--5) & 0.4000 & 0.6000 & 0.6000 & 0.6000 & 0.0000 \\
GPT-4o & Reactor Physics (Q6--10) & 0.7000 & 1.0000 & 0.4000 & 0.6000 & 0.0000 \\
GPT-4o & Thermal Shield (Q11--15) & 0.8000 & 1.0000 & 1.0000 & 0.2500 & 0.5300 \\
GPT-4o & \textbf{Average (15Q)} & \textbf{0.6333} & \textbf{0.8667} & \textbf{0.6667} & \textbf{0.4833} & \textbf{0.1767} \\
\bottomrule
\end{tabular}
\end{table*}

Representative failure modes from the page level benchmark are summarized in Table~\ref{tab:failure_modes}. These examples illustrate how distinct visual grounding errors propagate into CoP, ViR, and HR. Although the averages in Table~\ref{tab:per_case_averages} capture overall trends, inspection at the query level revealed different failure mechanisms. In Q1 (parabolic gate), some models conflated plotted coordinate axes with physical dimensions or fabricated unlabeled lengths (``2 ft'', ``4 ft''), leading to reduced numeric precision ($\text{CoP}_N$) and elevated HR. In Q6 and Q10 (reactor physics schematic), the ``GPT-4.x'' family either omitted explicitly labeled axial markers or asserted that no axial dependence was depicted despite clear visual cues, resulting in diminished ViR and partial CoP. In Q15 (thermal shield modeling), GPT-4o applied an incorrect geometric abstraction (cylindrical shell rather than slab) even when radii were correctly recovered, indicating that failures can arise at the level of physical interpretation rather than raw visual extraction. By contrast, derived inferences consistent with both figure and text (e.g., shield thickness inferred from labeled radii in Q11 by GPT-5) did not degrade any metric and were not penalized.

\begin{table}[htbp]
\centering
\small
\caption{Representative failure modes observed in the page level benchmark. The table summarizes characteristic error patterns across models and queries and highlights how different failure mechanisms impact CoP, ViR, and HR.}
\label{tab:failure_modes}
\resizebox{\textwidth}{!}{%
\begin{tabular}{|c|c|c|p{3.5cm}|c|p{7.2cm}|}
\hline
\textbf{Query} &
\textbf{Case} &
\textbf{Model} &
\textbf{Observed Failure} &
\textbf{Impacted Metrics} &
\textbf{Model Summary} \\
\hline

Q1 &
Calculus &
GPT-4o &
Visual omission  &
CoP $\downarrow$, ViR $\downarrow$ &
The model reported that the figure contains no geometric information, despite being explicitly labeled with dimensions ($10 ft$ and $6 ft$). \\
\hline

Q1 &
Calculus &
GPT-4.1 &
Axis vs.\ dimension confusion &
CoP$_N$ $\downarrow$, HR $\uparrow$ &
Interprets plotted coordinate axes as physical dimensions and fabricates a width of ``4''. \\
\hline

Q1 &
Calculus &
GPT-5.1 &
Partial numeric hallucination &
CoP$_N$ $\downarrow$, HR $\uparrow$ &
Correctly identifies schematic intent but fabricates vertical dimensions (``2 ft'', ``4 ft''). \\
\hline

Q6 &
Reactor Physics &
GPT-4o &
Figure not recognized &
CoP $\downarrow$, ViR $\downarrow$ &
Claims Figure~3--3 provides no spatial or coordinate information despite clearly labeled geometry. \\
\hline

Q10 &
Reactor Physics &
GPT-5.1 &
Axial omission &
ViR $\downarrow$, CoP$_S$ $\downarrow$ &
Correctly interprets radial profiles but incorrectly states that axial dependence is absent. \\
\hline

Q15 &
Thermal Shield &
GPT-4o &
Geometry abstraction error &
CoP$_S$ $\downarrow$ &
Models the thermal shield as a cylindrical shell instead of a slab, contradicting explicit textual assumptions. \\
\hline

Q11 &
Thermal Shield &
GPT-5 &
Correct derived inference (not penalized) &
N/A &
Correctly infers shield thickness from labeled radii ($R_2 - R_1$), consistent with figure semantics. \\
\hline

\end{tabular}
}
\end{table}

\subsubsection{Cross Model Knowledge Base Effects}

\begin{figure*}[htbp]
    \centering
    \begin{subfigure}[t]{0.32\textwidth}
        \centering
        \includegraphics[width=\linewidth]{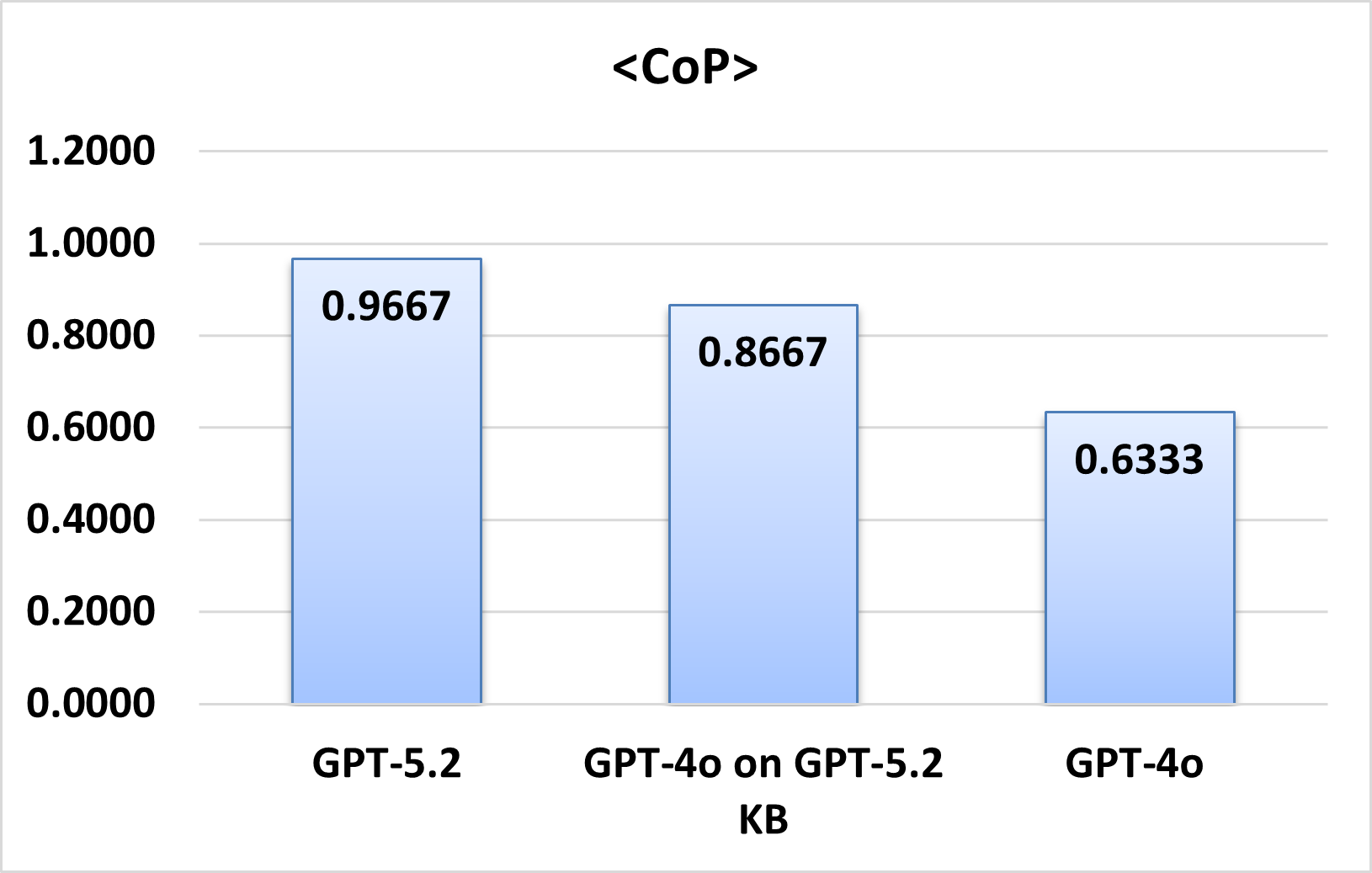}
        \caption{CoP}
    \end{subfigure}
    \hfill
    \begin{subfigure}[t]{0.32\textwidth}
        \centering
        \includegraphics[width=\linewidth]{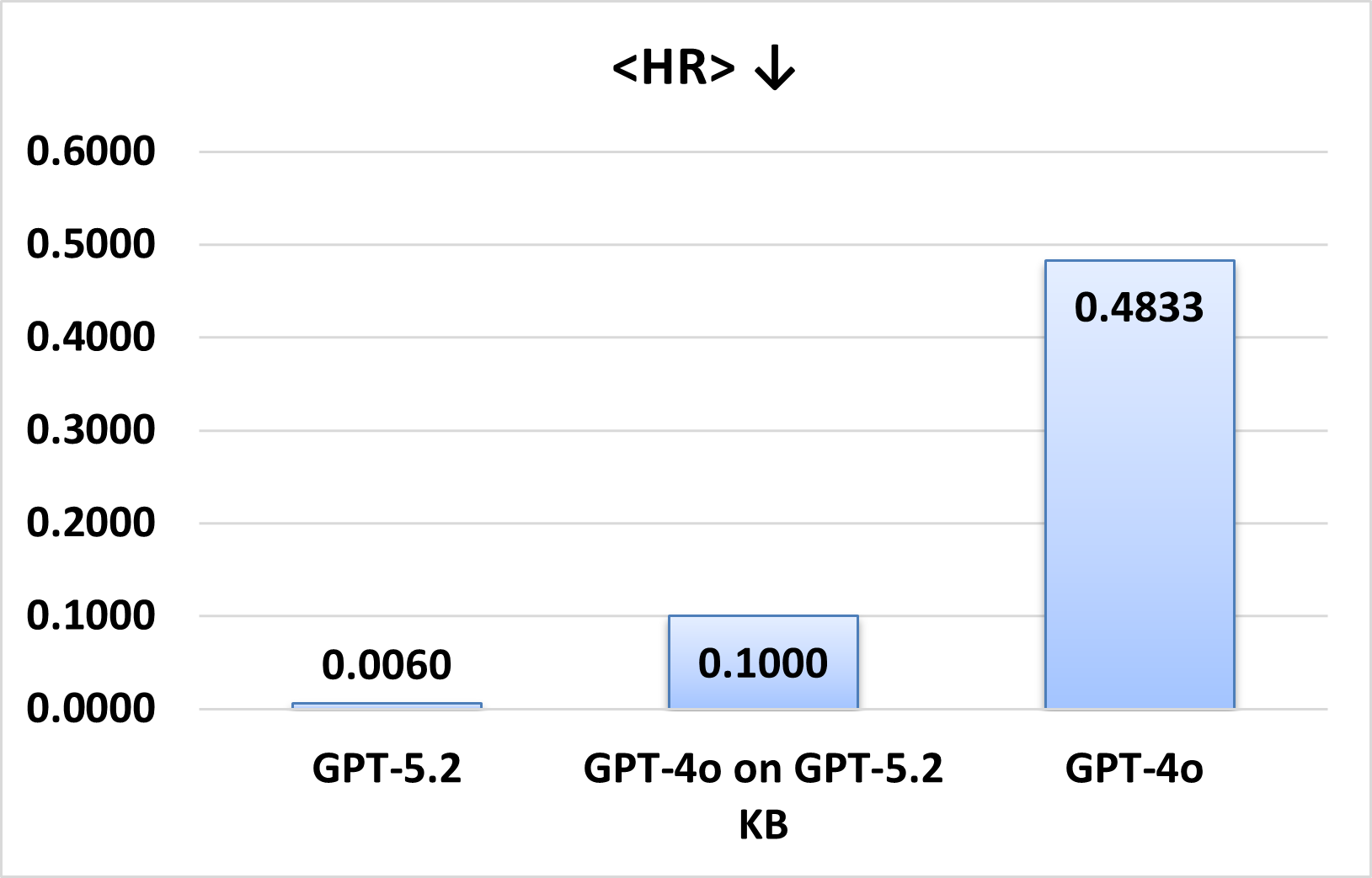}
        \caption{HR}
    \end{subfigure}
    \hfill
    \begin{subfigure}[t]{0.32\textwidth}
        \centering
        \includegraphics[width=\linewidth]{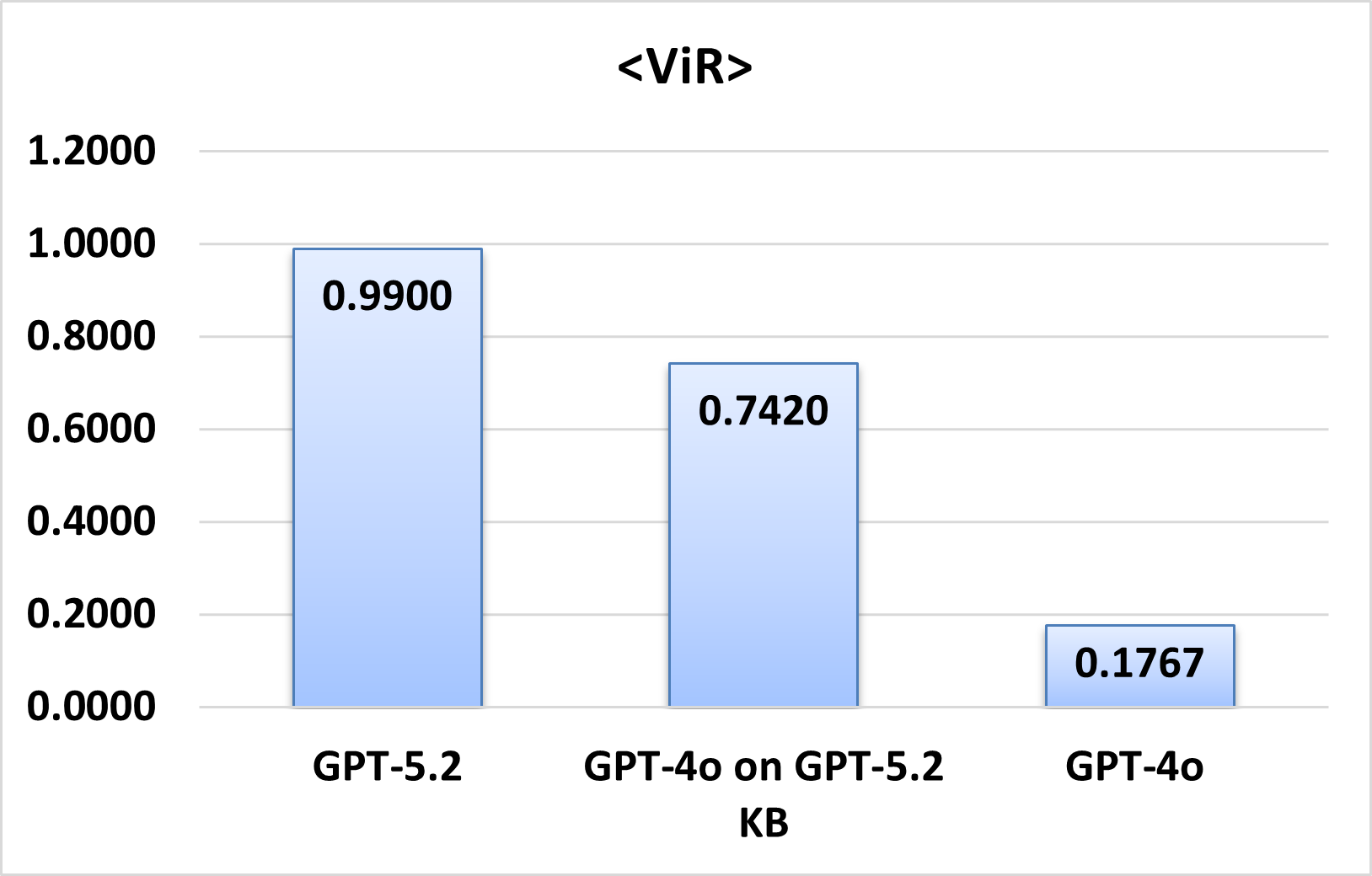}
        \caption{ViR}
    \end{subfigure}

    \caption{Effect of knowledge base fidelity on downstream model performance. Page level benchmark results averaged across all 15 queries, comparing GPT-5.2 queried on its own KB, GPT-4o queried on a GPT-5.2 built KB, and GPT-4o queried on its own KB. Querying GPT-4o on a higher fidelity multimodal KB improves CoP and ViR and reduces HR.}
    \label{fig:cross_model_kb_effect}
\end{figure*}

When GPT-4o (the lowest performing model in the page level benchmark) was queried against a knowledge base constructed using the higher capability VLM configuration (GPT-5.2), answer quality improved substantially. As shown in Figure~\ref{fig:cross_model_kb_effect}, GPT-4o's $\overline{\mathrm{CoP}}$ increased from 0.633 to 0.867, $\overline{\mathrm{HR}}$ decreased from 0.483 to 0.100, and $\overline{\mathrm{ViR}}$ improved from 0.177 to 0.742 relative to querying its own page level KB. In this configuration, GPT-4o approached or exceeded the CoP and ViR achieved by intermediate models (GPT-5.1 and GPT-4.1), despite its lower intrinsic reasoning capacity. Qualitatively, these gains were most evident in queries requiring precise interpretation of visual semantics and geometric assumptions (e.g., Q1, Q6, and Q15), as summarized in Table~\ref{tab:cross_model_improvements}.

\begin{table}[H]
\centering
\small
\caption{Representative query level improvements when GPT-4o queried a GPT-5.2 constructed knowledge base versus its own page level KB. Higher fidelity visual parsing during KB construction improved answer correctness, visual grounding, and hallucination suppression for a lower capability reasoning model.}
\label{tab:cross_model_improvements}
\resizebox{\textwidth}{!}{%
\begin{tabular}{|c|p{1.5cm}|p{5cm}|p{4.6cm}|p{4.6cm}|}
\hline
\textbf{Query} &
\textbf{Case} &
\textbf{Baseline: (GPT-4o on its own KB)} &
\textbf{GPT-4o on GPT-5.2 KB} &
\textbf{Observed Improvement Mechanism} \\
\hline

Q1 &
Calculus &
Confused plotted axes with physical dimensions; fabricated unlabeled distances (e.g., ``4 ft''), reducing CoP and increasing HR. &
Correctly identifies labeled dimensions (10 ft, 6 ft) as physical lengths and explicitly rejects axis interpretation; no fabricated values. &
High fidelity visual fact extraction during KB construction eliminated ambiguity between coordinate axes and physical dimensions, reducing hallucination and restoring numeric precision. \\
\hline

Q6 &
Reactor Physics &
Claims figure provides no spatial or coordinate information; omits axial markers and extrapolation lengths. &
Correctly identifies radial and axial directions, extrapolated boundaries, and cylindrical geometry using figure labels. &
Dense visual annotations (e.g., $r=0$, $R$, $R_e$, $\pm L/2$, $\pm L_e/2$) preserved in the KB enabled accurate visual grounding without requiring stronger model reasoning. \\
\hline

Q15 &
Thermal Shield &
Applies incorrect geometric abstraction (cylindrical shell) despite textual assumptions, reducing CoP. &
Correctly treats the thermal shield as a slab and the core as an infinite plane source, consistent with both text and figure. &
Explicit encoding of geometric assumptions during ingestion reduced reliance on implicit reasoning, preventing higher level modeling errors. \\
\hline

\end{tabular}
}
\end{table}

These page level experiments show that RAG performance depends on both ingestion fidelity and model reasoning. In this study, GPT-5.2 produced the most reliable VLM outputs during KB construction, and even weaker reasoning models improved when they queried a KB built with this higher fidelity parser. Based on these results, GPT-5.2 was selected as the VLM configuration for subsequent document level experiments and for constructing the persistent 3S knowledge bases used in this study.

\subsection{Scope of RADIANT-LLM Relative to General Purpose Deployments}
\label{subsec:scope_vs_commercial}

Before examining retrieval sensitivity under context scaling, we compared RADIANT-LLM against several GPT-5.2 deployments on the same UNFSF 3S benchmark described in Section~\ref{subsec:Context_Expansion}. We evaluated three general purpose deployments based on GPT-5.2: (i) ChatGPT powered by GPT-5.2 instant with automatic web access, (ii)Texas A\&M Chat (GPT-5.2 with institutional access but no web access), and (iii) a GPT-5.2 frozen model with no external context. Their results were compared to RADIANT-LLM powered by GPT-5, queried on the 250 source local knowledge base constructed using GPT-5.2 as the VLM. This configuration follows the earlier observation in Section~\ref{subsec:page_level_sensitivity} that a weaker reasoning model can query a higher fidelity multimodal KB created with a stronger VLM.

This comparison is not intended as a head to head competition. Instead, it probes practical limitations of general purpose LLM deployments under constraints relevant to nuclear safety, security, and safeguards analysis. Commercial systems evolve rapidly, so future model updates may exhibit different performance characteristics than those reported here.

\begin{figure}[htbp]
\centering
\includegraphics[width=0.95\linewidth]{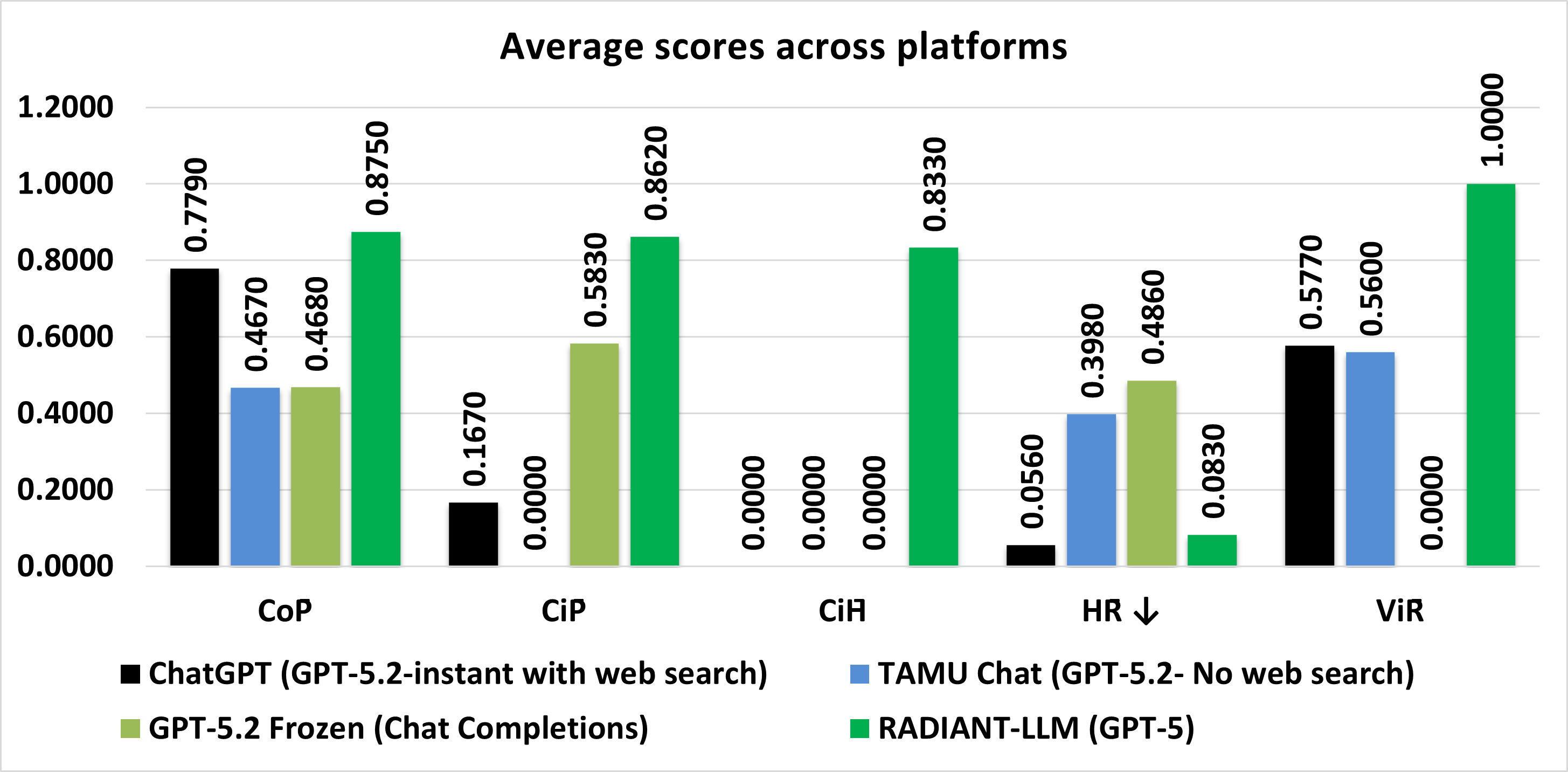}
\caption{Average CoP, CiP, CiH, HR (lower is better), and ViR across UNFSF queries for different GPT-5.2 deployments.}
\label{fig:average_scores}
\end{figure}

As shown in Figure~\ref{fig:average_scores}, RADIANT-LLM (with GPT-5) achieved the highest overall correctness (CoP = 0.875), perfect visual recall (ViR = 1.000), a low hallucination rate (HR = 0.083), and strong citation performance (CiP = 0.862, CiH = 0.833) across the twelve UNFSF benchmark queries. In contrast, ChatGPT with web access reached a moderate CoP (0.779) and a relatively low HR (0.056). It provided some citations to general nuclear and regulatory sources (CiP = 0.167), but none of its citations hit the expert defined anchors at the page or figure level (CiH = 0). Visual fact recovery was also partial (ViR = 0.577). The Texas A\&M Chat deployment (GPT-5.2 with no web search or external retrieval) exhibited the lowest overall correctness (CoP = 0.467), elevated hallucination (HR = 0.398), and no valid citations (CiP = CiH = 0), despite moderate visual recall on the limited figure based questions (ViR = 0.560). The frozen GPT-5.2 model, queried without any external context, showed similar CoP (0.468) but the highest hallucination rate among the evaluated systems (HR = 0.486). It achieved a moderate CiP score (0.583) by citing general regulatory or IAEA style documents, but it failed to hit any of the specific UNFSF pages or figures used to define the benchmark (CiH = 0) and exhibited no visual fact recovery (ViR = 0.000).

The similar CoP obtained for the frozen GPT-5.2 model and the institutional GPT-5.2 deployment without external retrieval suggests that, in this benchmark, an API model without retrieval behaves much like a frozen engine with no external context. Achieving reliable performance for nuclear 3S workflows therefore requires explicit retrieval, tool orchestration, and multimodal RAG augmentation of the kind implemented in this study.

Note that by definition, CiP and CiH play complementary roles. CiP reflects whether the citations, even outside the corpus, lead to sources that support the model's answer. This is why ChatGPT and the frozen GPT-5.2 model received nonzero CiP when they pointed to broadly relevant standards or guidance, even if those sources were outside the curated 3S corpus. CiH, by contrast, is stricter and only rewards citations that hit the expert defined anchors $E_i^*$ at the page, section, or figure level. As a result, all general purpose deployments showed CiH = 0 because none of their citations landed on the benchmarked UNFSF document locations, while RADIANT-LLM achieves high CiP and CiH by repeatedly citing the correct pages and figures inside its local knowledge base.

Overall, these results highlight a limitation of general purpose LLM deployments for nuclear 3S applications. Even with web access, they do not provide built in control over which documents are used, how figures are represented, or how precisely citations are anchored to pages and figures. In contrast, RADIANT-LLM constructs a persistent, structured, and locally controlled multimodal knowledge base from domain technical documents, with explicit page and figure identifiers and associated metadata. This design supports reproducible and auditable question answering with traceable citations for repeated 3S workflows. RADIANT-LLM should therefore be viewed as complementary to commercial LLM platforms, providing a dedicated multimodal RAG layer that enables secure knowledge ingestion, precise citation, and physics aware visual reasoning beyond one off PDF question answering.

\begin{table}[H]
\centering
\small
\caption{Representative per query failure modes explaining metric degradation in general purpose GPT-5.2 deployments on the UNFSF benchmark. Each case highlights confident but incorrect behavior or citation drift observed in the YAML scoring files.}
\label{tab:platform_diagnostic_failures}
\resizebox{\textwidth}{!}{%
\begin{tabular}{|c|c|p{4.2cm}|p{4.8cm}|p{4.2cm}|}
\hline
\textbf{Query} &
\textbf{Model} &
\textbf{Observed Behavior} &
\textbf{Failure Mode} &
\textbf{Metric Impact} \\
\hline

Q9 &
TAMU Chat &
Correctly states Pu-239 detection mass but assigns incorrect shielding material and confidence level; omits false alarm constraint. &
Confident numeric substitution inconsistent with UNFSF Section~3.3.2 requirements. &
CoP$\downarrow$, HR$\uparrow$ \\
\hline

Q11 &
TAMU Chat &
Claims detection of 10 g U-233 within $\sim$1 s under lead shielding. &
Severe numeric hallucination contradicting UNFSF thresholds by orders of magnitude. &
CoP$\downarrow$, HR$\uparrow$ \\
\hline

Q12 &
TAMU Chat &
Provides precise metal and explosive detector thresholds not present in UNFSF text. &
Fabrication of plausible but unsupported detector performance values. &
CoP$\downarrow$, HR$\uparrow$ \\
\hline

Q6 &
GPT-5.2 Frozen &
Identifies surveillance/CCTV as the selected dual use system. &
Incorrect system selection; UNFSF specifies an exit/entry control system (Section~3.2). &
CoP$\downarrow$, HR$\uparrow$ \\
\hline

Q4 to Q8 &
GPT-5.2 Frozen &
Cites NRC and IAEA regulations accurately but fails to recover UNFSF design logic. &
Source drift: correct external citations without document specific grounding. &
CiP$\uparrow$, CiH$=0$ \\
\hline

Q1 to Q3 &
ChatGPT (web) &
Explains 3S concepts correctly but fails to recover figure specific overlap labels. &
Partial visual grounding without figure level semantic recovery. &
ViR$\downarrow$, CiH$=0$ \\
\hline

\end{tabular}
}
\end{table}

Inspection of the per query YAML scoring files in Table~\ref{tab:platform_diagnostic_failures} suggests that the main degradation in general purpose deployments arises from confident but incorrect numerical answers and source drift. For example, in the detector system queries (Q9 to Q12) for SNM doorway monitors, both TAMU Chat and the frozen GPT-5.2 model frequently asserted precise performance thresholds (e.g., detection mass, confidence level, false alarm rate) that were inconsistent with the UNFSF specifications, leading to sharp drops in $CoP_N$, hence $CoP$ (Equation~\eqref{eq:cp_def}), and elevated hallucination rates. In the regulatory and framework questions (Q4 to Q8), the frozen model often cited valid NRC or IAEA documents, yielding nonzero CiP, but failed to recover the UNFSF design logic and anchors, resulting in CiH = 0. These cases explain why average correctness and hallucination metrics degrade even when answers appear well structured and authoritative.
\subsection{Effects of Context Scaling on Retrieval Sensitivity}
\label{subsec:Context_Expansion}

For the context expansion benchmark, all queries were drawn from \textit{``Application of Framework for Integrating Safety, Security and Safeguards (3Ss) into the Design of Used Nuclear Fuel Storage Facility (LA-UR-14-27045)''}~\cite{badwan2015application}. The benchmark questions were designed to reflect both conceptual and implementation aspects of 3S integration. Queries Q1 to Q3 were explicitly grounded in the 3S interface schematic already introduced in Figure~\ref{fig:3s_diagram}. These queries probed how safety, security, and safeguards shared operational functions, how their roles and primary objectives were distinguished, and how dual use systems supporting both safeguards and security could be represented in schematics. The remaining queries (Q4 to Q12) targeted the surrounding text and design requirements, covering (i) the regulatory frameworks governing physical protection and material control and accounting (MC\&A), (ii) performance objectives and justification criteria for dual use systems, (iii) the defined application scope of the integration framework, and (iv) quantitative detection and false alarm thresholds for spent nuclear material (SNM) doorway monitors and metal or explosive detectors in the used nuclear fuel storage facility (UNFSF). Together, these questions provided a 3S focused test and enabled a structured evaluation of retrieval fidelity (CiP, CiH), hallucination behavior (HR), and answer correctness (CoP) under expanding knowledge bases.

\begin{table}[htbp]
\centering
\caption{Average performance metrics for GPT-5.2 evaluated on the UNFSF visual-RAG benchmark as a function of retrieved context size. }
\label{tab:gpt52_context_scaling}
\begin{tabular}{lcccccc}
\hline
Context level &
$\overline{\text{CoP}}$ &
$\overline{\text{CiP}}$ &
$\overline{\text{CiH}}$ &
$\overline{\text{HR}}$ &
$\overline{\text{ViR}}$ \\
\hline
1 source (27p)   & 0.979 & 0.848 & 0.583 & 0.000 & 1.000 \\
5 sources (265p) & 0.946 & 1.000 & 1.000 & 0.036 & 0.780 \\
10 sources (408p) & 0.946 & 0.958 & 0.917 & 0.031 & 0.953 \\
20 sources (826p) & 0.967 & 1.000 & 0.917 & 0.014 & 1.000 \\
50 sources (2004p) & 0.849 & 1.000 & 0.667 & 0.163 & 0.952 \\
100 sources (6349p, R1) & 0.599 & 0.708 & 0.750 & 0.467 & 0.083 \\
100 sources (6349p, R2) & 0.938 & 0.917 & 1.000 & 0.000 & 1.000 \\
150 sources (9392p) & 0.883 & 0.917 & 0.917 & 0.073 & 0.905 \\
200 sources (12406p) & 0.925 & 0.775 & 0.750 & 0.177 & 0.905 \\
250 sources (16141p) & 0.863 & 0.900 & 0.833 & 0.083 & 1.000 \\
\hline
\end{tabular}
\end{table}

\begin{figure}[htbp]
\centering
\begin{subfigure}[b]{0.48\linewidth}
  \centering
  \includegraphics[width=\linewidth]{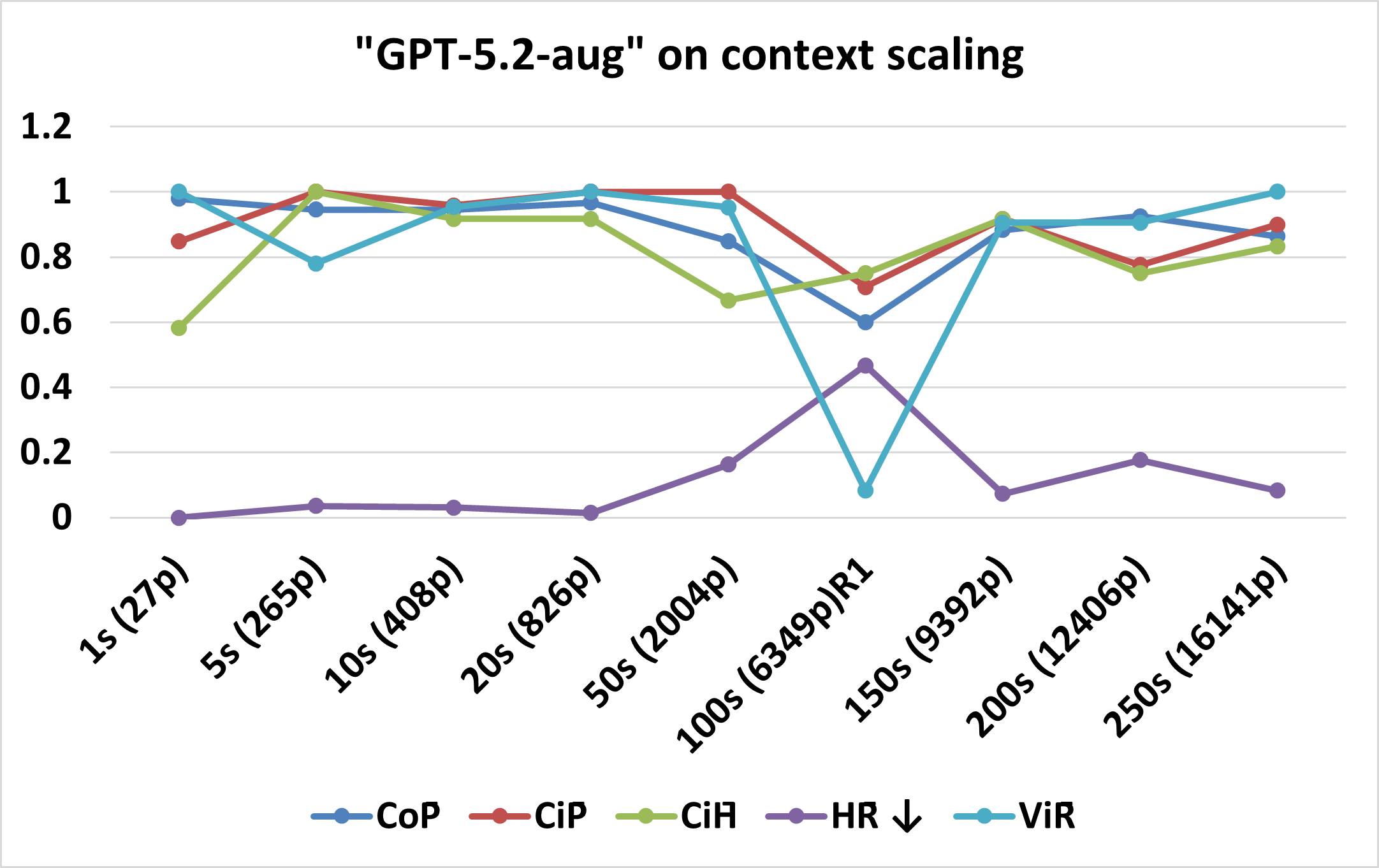}
  \caption{GPT-5.2, all context levels (R1)}
  \label{fig:gpt52_context_scaling_a}
\end{subfigure}
\hfill
\begin{subfigure}[b]{0.48\linewidth}
  \centering
  \includegraphics[width=\linewidth]{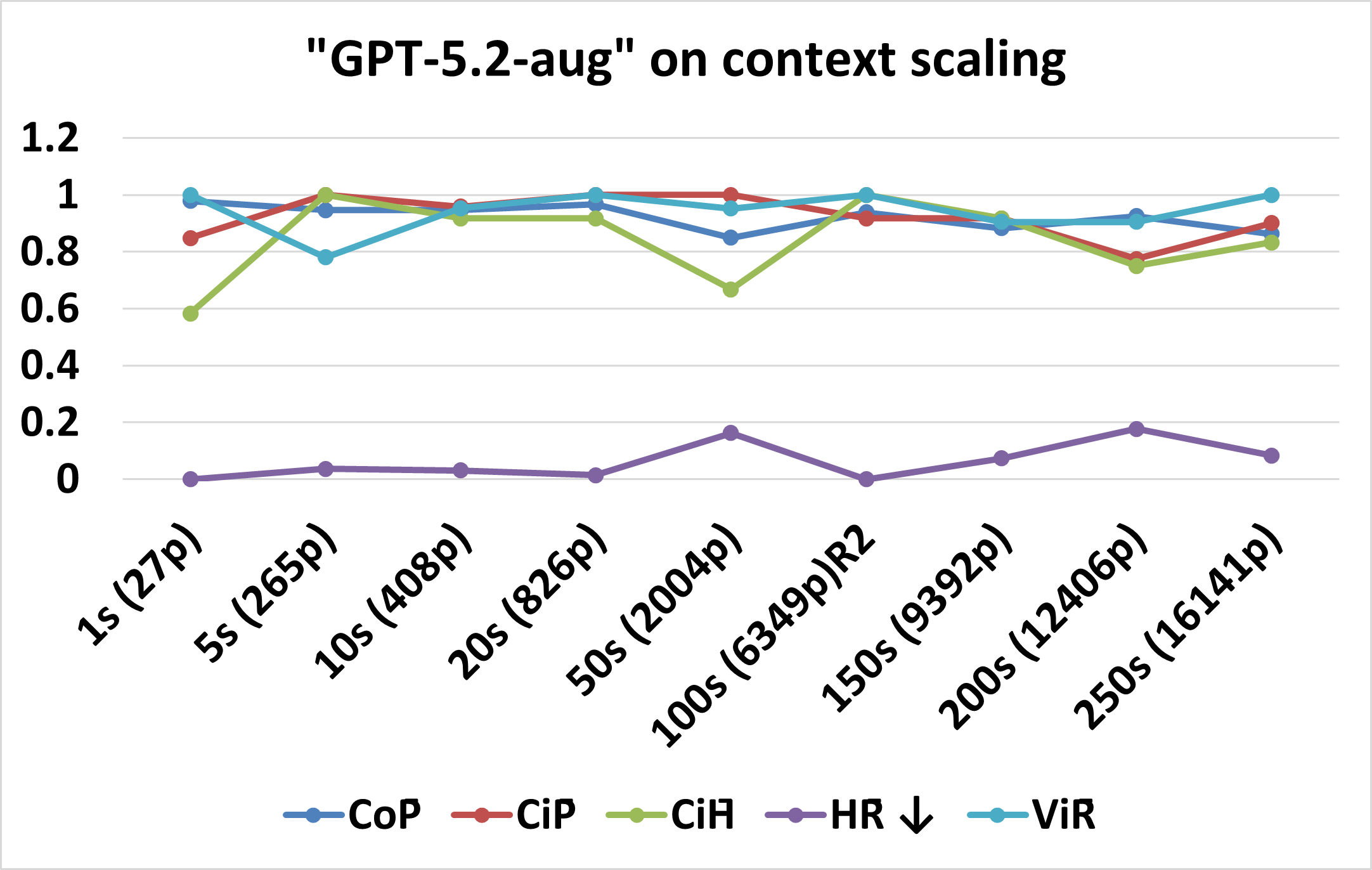}
  \caption{GPT-5.2, repeat at 100 sources (R2)}
  \label{fig:gpt52_context_scaling_b}
\end{subfigure}

\caption{Context scaling for GPT-5.2 on the UNFSF Visual-RAG benchmark. Shown are average CoP, CiP, CiH, HR (lower is better), and ViR as a function of retrieved context size. Panel (a) includes all context levels from run~1 and highlights a degraded regime at 100 sources. Panel (b) shows a repeat of the 100 source setting (run~2), where performance returned to the typical high performance band.}
\label{fig:gpt52_context_scaling}
\end{figure}

Table~\ref{tab:gpt52_context_scaling} and Figure~\ref{fig:gpt52_context_scaling} summarize GPT-5.2 performance on this UNFSF Visual-RAG benchmark as the number of retrieved sources increased from 1 (27 pages) to about 250 (16{,}141 pages). Across most context levels, $\overline{\text{CoP}}$ remained in a high band (typically $\gtrsim 0.85$), $\overline{\text{ViR}}$ stayed close to unity for visual queries, $\overline{\text{CiP}}$ and $\overline{\text{CiH}}$ were generally high, and $\overline{\text{HR}}$ remained low. At small to moderate context sizes (1 to 20 sources), adding surrounding text often improved or preserved CoP and ViR while keeping HR near zero.


At larger context sizes (50 to 250 sources), performance continued to lie within this high band, but one 100 source experiment (run~1) exhibited a marked degradation: CoP and ViR dropped, while HR increased, indicating a transient failure regime where dense, overlapping evidence appeared to exacerbate anchor competition and unsupported inference. A second 100 source experiment (run~2), performed under identical settings, recovered to near peak performance ($\overline{\text{CoP}}=0.938$, $\overline{\text{ViR}}=1.000$, $\overline{\text{HR}}=0.000$, $\overline{\text{CiP}}=0.917$, $\overline{\text{CiH}}=1.000$), and the 150 to 250 source regimes also maintained high CoP and ViR with modest HR.

This pattern is consistent with the probabilistic, autoregressive nature of LLM generation discussed earlier in Section~\ref{sec:background}. As the number of retrieved passages increases, the model's response distribution can become more sensitive to retrieval noise and small changes in internal sampling, leading to occasional but recoverable failure regimes rather than systematic collapse. Representative diagnostic questions in Table~\ref{tab:context_failure_modes} illustrate how these regimes manifest in practice (e.g., omitted false alarm constraints, mixed safeguards and security interfaces, or missing central 3S overlap regions at 100 sources in run~1).

\begin{table}[htbp]
\centering
\scriptsize
\caption{Representative diagnostic questions illustrating failure modes and recovery trends across context scaling regimes. Each row highlights the question focus, observed model behavior, and impacted metrics.}
\label{tab:context_failure_modes}

\newcommand{\colQ}{0.07\linewidth}   
\newcommand{\colFocus}{0.25\linewidth} 
\newcommand{\colCtx}{0.12\linewidth} 
\newcommand{\colObs}{0.28\linewidth} 
\newcommand{\colMet}{0.18\linewidth} 

\begin{tabular}{
p{\colQ}
p{\colFocus}
p{\colCtx}
p{\colObs}
p{\colMet}
}
\hline
\textbf{Q} & \textbf{Question focus} & \textbf{Context regime} & \textbf{Observed behavior} & \textbf{Impacted\newline metrics} \\
\hline
Q1 & 3S shared functions (figure) 
& 100 (R1) 
& Central overlap omitted 
& ViR$\downarrow$, HR$\uparrow$ \\

Q2 & Domain objectives (figure) 
& 100 (R1) 
& Role labels partially inferred 
& ViR$\downarrow$ \\

Q3 & Safeguards and security interface 
& 100 (R1) 
& Interface functions mixed 
& HR$\uparrow$ \\

Q9 & Pu-239 detection requirements 
& 100 (R1) 
& False alarm constraint omitted 
& CoP$\downarrow$ \\

Q10 & SNM false alarm limits 
& 100 (R1) 
& Isotope coverage incomplete 
& CoP$\downarrow$ \\

Q7 & Dual use system justification 
& 10 to 50 
& Extra contextual citations introduced 
& CiH$\downarrow$ \\

Q8 & Framework application scope 
& 1 to 5 
& Explicit ``not in KB'' response 
& HR$\downarrow$ \\
\hline
\end{tabular}
\end{table}
Overall, these results show that, for this 3S focused UNFSF benchmark, Visual-RAG with GPT-5.2 maintained strong factual accuracy, citation anchoring, and visual grounding across a wide range of context sizes. Performance did not degrade monotonically with context; however, dense and overlapping evidence can occasionally trigger poor retrieval or generation paths, as seen in the 100 source run~1. In practical 3S workflows, this motivates pairing high fidelity multimodal ingestion with retrieval and supervision strategies that monitor CoP, HR, CiP, CiH, and ViR and enable mitigation steps such as retrieval tuning, reranking, or targeted human oversight. These context scaling trends are consistent with the cross platform comparison in Section~\ref{subsec:scope_vs_commercial}, which showed that a dedicated, locally controlled multimodal KB is important for maintaining high CoP, CiP, CiH, and ViR relative to general purpose GPT-5.2 deployments.

\section{Conclusions}
\label{sec:Conclusion}

This work presented RADIANT-LLM, a multi-modal retrieval-augmented generation (RAG) framework for reliable, traceable decision support in safety-critical nuclear engineering. By combining a multi-modal ingestion pipeline, a JSON-based local knowledge base, and domain-aware evaluation metrics (CoP, CiP, CiH, HR, and ViR), the framework transforms the use of technical nuclear documents from a manual, error-prone process into a document-cited, visually grounded workflow aligned with the reliability and auditability requirements of nuclear Safety, Security, and Safeguards (3S) applications.

Quantitatively, when backed by a high-fidelity multi-modal KB constructed with a strong VLM (GPT-5.2), even a weaker reasoning model (GPT-5) achieved high CoP, low HR, near-perfect ViR, and strong citation behavior across both page-level and context-expansion benchmarks. On the UNFSF benchmark, CoP, CiP, CiH, and ViR generally remained above 0.85 from 1 to 250 retrieved sources despite increasing document density. An occasional dip at 100 sources in one run, followed by recovery in a repeat, reflects the stochastic nature of autoregressive generation rather than systematic degradation. The stability of these metrics across the remaining runs demonstrates the robustness of the methodology under dense, potentially noisy evidence.

Comparison against GPT-5.2-based deployments outside the RADIANT-LLM architecture (ChatGPT with web access, TAMU Chat, and a frozen GPT-5.2 model) confirmed that, even for state-of-the-art foundation models, the absence of a dedicated local multi-modal KB leads to weaker CoP, higher hallucination rates, and misaligned citations, particularly for corpus-specific nuclear documents. This is consistent with findings in the reliability engineering literature, where domain-grounded RAG frameworks have outperformed general-purpose deployments in fault diagnosis~\cite{ZHENG2024110382}, accident analysis~\cite{ZHANG2026112333}, human reliability assessment~\cite{XIAO2026111585}, and infrastructure maintenance~\cite{YU2026111973}. RADIANT-LLM extends this line of work by unifying text, figures, schematics, and regulatory documents within a single multi-modal retrieval and provenance-enforcement framework.

The architecture is designed for broader applicability beyond the benchmarks reported here. Its JSON-based, page- and figure-resolved KB structure is portable and version-controlled, suitable for facility-specific KBs (SMRs, Gen-IV reactors, fuel-cycle facilities), and the claim- and citation-level metrics are model-agnostic and reusable across future LLM/VLM configurations.

Several directions could strengthen the framework: scaling to larger corpora (IAEA, NRC, design certification submissions), benchmarking additional LLM backends to clarify cost-reliability trade-offs, tighter coupling with M\&S frameworks and digital twins for risk-informed decision support, and integration with structured risk analysis methodologies such as functional resonance analysis~\cite{LIANG2026112416} or knowledge-graph-based human reliability assessment~\cite{XIAO2026111585}. User studies and deployment pilots in operational 3S workflows remain a priority. Because the architecture prioritizes local control and traceable evaluation over any specific underlying model, it can absorb future improvements in commercial LLMs and VLMs, providing a durable foundation for reliable AI-assisted decision support in nuclear engineering and a transferable template for other safety-critical domains.

\section*{Acknowledgements}
This work is primarily supported by the U.S. Department of Energy Office of Nuclear Energy Distinguished Early Career Program under contract number DE-NE0009468.

Additional support is provided by the Texas A\&M Institute of Data Science (TAMIDS) Seed Program for AI, Computing, and Data Science.


\bibliographystyle{elsarticle-num}
\bibliography{references}

@article{Pandya2023AutomatingOrganizations,
  title = {{Automating Customer Service using LangChain: Building custom open-source GPT Chatbot for organizations}},
  year = {2023},
  journal = {arXiv preprint arXiv:2310.05421},
  author = {Pandya, Keivalya and Holia, Mehfuza},
}

@article{Jeong2023GenerativeFramework,
  title = {{Generative AI service implementation using LLM application architecture: based on RAG model and LangChain framework}},
  year = {2023},
  journal = {Journal of Intelligence and Information Systems},
  author = {Jeong, Cheonsu},
  pages = {129--164},
  volume = {29},
  publisher = {Korea Intelligent Information System Society},
}

@misc{H.Chase2023GitHubApplications,
  title = {{GitHub - langchain-ai/langchain: Build context-aware reasoning applications}},
  year = {2023},
  author = {{H. Chase}},
}

@article{blecher2023nougat,
  title = {Nougat: Neural optical understanding for academic documents},
  author = {Blecher, Lukas and Cucurull, Guillem and Scialom, Thomas and Stojnic, Robert},
  journal = {arXiv preprint arXiv:2308.13418},
  year = {2023},
}

@inproceedings{Singh2024RevolutionizingModel,
  title = {{Revolutionizing mental health care through langchain: A journey with a large language model}},
  year = {2024},
  booktitle = {2024 IEEE 14th Annual Computing and Communication Workshop and Conference (CCWC)},
  author = {Singh, Aditi and Ehtesham, Abul and Mahmud, Saifuddin and Kim, Jong-Hoon},
  pages = {73--78},
}

@article{kwon2024sentiment,
  title = {Sentiment analysis of the United States public support of nuclear power on social media using large language models},
  author = {Kwon, O Hwang and Vu, Katie and Bhargava, Naman and Radaideh, Mohammed I and Cooper, Jacob and Joynt, Veda and Radaideh, Majdi I},
  journal = {Renewable and Sustainable Energy Reviews},
  volume = {200},
  pages = {114570},
  year = {2024},
  publisher = {Elsevier},
}

@inproceedings{fan2024survey,
  title = {A survey on rag meeting llms: Towards retrieval-augmented large language models},
  author = {Fan, Wenqi and Ding, Yujuan and Ning, Liangbo and Wang, Shijie and Li, Hengyun and Yin, Dawei and Chua, Tat-Seng and Li, Qing},
  booktitle = {Proceedings of the 30th ACM SIGKDD Conference on Knowledge Discovery and Data Mining},
  pages = {6491--6501},
  year = {2024},
}

@article{suresh2024towards,
  title = {Towards a RAG-based summarization for the Electron Ion Collider},
  author = {Suresh, Karthik and Kackar, Neeltje and Schleck, Luke and Fanelli, Cristiano},
  journal = {Journal of Instrumentation},
  volume = {19},
  pages = {C07006},
  year = {2024},
  publisher = {IOP Publishing},
}

@phdthesis{iob2024nuclear,
  title = {Nuclear Security: A Natural Language Processing Generative Approach},
  author = {Iob, Gabriele},
  year = {2024},
  school = {Politecnico di Torino},
}

@article{turner2016virtual,
  title = {The virtual environment for reactor applications (VERA): design and architecture},
  author = {Turner, John A and Clarno, Kevin and Sieger, Matt and Bartlett, Roscoe and Collins, Benjamin and Pawlowski, Roger and Schmidt, Rodney and Summers, Randall},
  journal = {Journal of Computational Physics},
  volume = {326},
  pages = {544--568},
  year = {2016},
  publisher = {Elsevier},
}

@article{turinsky2016modeling,
  title = {Modeling and simulation challenges pursued by the Consortium for Advanced Simulation of Light Water Reactors (CASL)},
  author = {Turinsky, Paul J and Kothe, Douglas B},
  journal = {Journal of Computational Physics},
  volume = {313},
  pages = {367--376},
  year = {2016},
  publisher = {Elsevier},
}

@inproceedings{godoy2020workflows,
  title = {Workflows Using Mantid},
  author = {Godoy, William F and Peterson, Peter F and Hahn, Steven E and Hetrick, John},
  booktitle = {Driving Scientific and Engineering Discoveries Through the Convergence of HPC, Big Data and AI: 17th Smoky Mountains Computational Sciences and Engineering Conference, SMC 2020, Oak Ridge, TN, USA, August 26-28, 2020, Revised Selected Papers},
  volume = {1315},
  pages = {175},
  year = {2020},
  organization = {Springer Nature},
}

@article{ndum2025automating,
  title = {Automating Monte Carlo simulations in nuclear engineering with domain knowledge-embedded large language model agents},
  author = {Ndum, Zavier Ndum and Tao, Jian and Ford, John and Liu, Yang},
  journal = {Energy and AI},
  pages = {100555},
  year = {2025},
  publisher = {Elsevier},
}

@article{annepaka2025large,
  title = {Large language models: a survey of their development, capabilities, and applications},
  author = {Annepaka, Yadagiri and Pakray, Partha},
  journal = {Knowledge and Information Systems},
  volume = {67},
  pages = {2967--3022},
  year = {2025},
  publisher = {Springer},
}

@article{Zhang2024ADiscovery,
  title = {{A Comprehensive Survey of Scientific Large Language Models and Their Applications in Scientific Discovery}},
  year = {2024},
  author = {Zhang, Yu and Chen, Xiusi and Jin, Bowen and Wang, Sheng and Ji, Shuiwang and Wang, Wei and Han, Jiawei},
}

@article{Zhao2023AModels,
  title = {A survey of large language models},
  author = {Zhao, Wayne Xin and Zhou, Kun and Li, Junyi and Tang, Tianyi and Wang, Xiaolei and Hou, Yupeng and Min, Yingqian and Zhang, Beichen and Zhang, Junjie and Dong, Zican and others},
  journal = {arXiv preprint arXiv:2303.18223},
  year = {2023},
}

@article{Blecher2023Nougat:Documents,
  title = {{Nougat: Neural Optical Understanding for Academic Documents}},
  year = {2023},
  author = {Blecher, Lukas and Cucurull, Guillem and Scialom, Thomas and Stojnic, Robert},
}

@book{Todreas1989NuclearFundamentals,
  title = {{Nuclear Systems I Thermal Hydraulic Fundamentals}},
  year = {1989},
  author = {Todreas, Neil E and Kazimi, Mujid S},
  address = {Boca Raton},
}

@article{diefenthaler2024ai,
  title = {AI-assisted detector design for the EIC (AID (2) E)},
  author = {Diefenthaler, M and Fanelli, C and Gerlach, LO and Guan, W and Horn, T and Jentsch, A and Lin, M and Nagai, K and Nayak, H and Pecar, C and others},
  journal = {Journal of Instrumentation},
  volume = {19},
  pages = {C07001},
  year = {2024},
  publisher = {IOP Publishing},
}

@article{xiao2024krail,
  title = {KRAIL: A Knowledge-Driven Framework for Base Human Reliability Analysis Integrating IDHEAS and Large Language Models},
  author = {Xiao, Xingyu and Chen, Peng and Qi, Ben and Zhao, Hongru and Liang, Jingang and Tong, Jiejuan and Wang, Haitao},
  journal = {arXiv preprint arXiv:2412.18627},
  year = {2024},
}

@article{gao2023retrieval,
  title = {Retrieval-augmented generation for large language models: A survey},
  author = {Gao, Yunfan and Xiong, Yun and Gao, Xinyu and Jia, Kangxiang and Pan, Jinliu and Bi, Yuxi and Dai, Yixin and Sun, Jiawei and Wang, Haofen and Wang, Haofen},
  journal = {arXiv preprint arXiv:2312.10997},
  volume = {2},
  year = {2023},
}

@article{wang2025dreams,
  title = {DREAMS: Density Functional Theory Based Research Engine for Agentic Materials Simulation},
  author = {Wang, Ziqi and Huang, Hongshuo and Zhao, Hancheng and Xu, Changwen and Zhu, Shang and Janssen, Jan and Viswanathan, Venkatasubramanian},
  journal = {arXiv preprint arXiv:2507.14267},
  year = {2025},
}

@article{roemer2024artificial,
  title = {Artificial intelligence model GPT4 narrowly fails simulated radiological protection exam},
  author = {Roemer, G and Li, A and Mahmood, U and Dauer, L and Bellamy, M},
  journal = {Journal of Radiological Protection},
  volume = {44},
  pages = {013502},
  year = {2024},
  publisher = {IOP Publishing},
}

@article{act2025regulation,
  title = {Regulation (EU) 2024/1689 of the European Parliament and of the Council. 2024},
  author = {Act, Artificial Intelligence},
  journal = {URL: https://eur-lex. europa. eu/eli/reg/2024/1689/oj/eng. Date of access},
  volume = {3},
  year = {2025},
}

@article{spelda2025security,
  title = {Security practices in AI development},
  author = {Spelda, Petr and Stritecky, Vit},
  journal = {AI \& SOCIETY},
  pages = {1--11},
  year = {2025},
  publisher = {Springer},
}

@article{taeihagh2025governance,
  title = {Governance of generative AI},
  author = {Taeihagh, Araz},
  journal = {Policy and society},
  volume = {44},
  pages = {1--22},
  year = {2025},
  publisher = {Oxford University Press UK},
}

@article{zaidan2024ai,
  title = {AI governance in a complex and rapidly changing regulatory landscape: A global perspective},
  author = {Zaidan, Esmat and Ibrahim, Imad Antoine},
  journal = {Humanities and Social Sciences Communications},
  volume = {11},
  year = {2024},
  publisher = {Springer Science and Business Media LLC},
}

@article{oumano2025comparison,
  title = {Comparison of Large Language Models’ Performance on 600 Nuclear Medicine Technology Board Examination--Style Questions},
  author = {Oumano, Michael A and Pickett, Shawn M},
  journal = {Journal of Nuclear Medicine Technology},
  volume = {24},
  year = {2025},
  pages = {269--335},
  publisher = {Society of Nuclear Medicine},
}

@article{KOO2025101632,
  title = {Advanced nuclear technologies in modern energy systems: A comparative risk assessment in Japan},
  journal = {Energy Strategy Reviews},
  volume = {57},
  pages = {101632},
  year = {2025},
  author = {Bonjun Koo and Kazuhiko Noguchi and Fumitaka Watanabe and Kotaro Kubo and Tadahiro Shibutani},
}

@article{smuha2021race,
  title = {From a ‘race to AI’to a ‘race to AI regulation’: regulatory competition for artificial intelligence},
  author = {Smuha, Nathalie A},
  journal = {Law, Innovation and Technology},
  volume = {13},
  pages = {57--84},
  year = {2021},
  publisher = {Taylor \& Francis},
}

@article{judge2025code,
  title = {When code isn’t law: rethinking regulation for artificial intelligence},
  author = {Judge, Brian and Nitzberg, Mark and Russell, Stuart},
  journal = {Policy and Society},
  volume = {44},
  pages = {85--97},
  year = {2025},
  publisher = {Oxford University Press UK},
}

@techreport{NIST.AI.800-1.ipd,
  title = {Managing Misuse Risk for Dual-Use Foundation Models},
  author = {U.S. AI Safety Institute},
  institution = {National Institute of Standards and Technology (NIST)},
  year = {2024},
  month = {July},
  type = {Initial Public Draft},
}

@article{zhang2025citation,
  title = {Citation accuracy challenges posed by large language models},
  author = {Zhang, Manlin and Zhao, Tianyu},
  journal = {JMIR Medical Education},
  volume = {11},
  pages = {e72998},
  year = {2025},
  publisher = {JMIR Publications Toronto, Canada},
}

@article{abouammoh2025perceptions,
  title = {Perceptions and earliest experiences of medical students and faculty with ChatGPT in medical education: qualitative study},
  author = {Abouammoh, Noura and Alhasan, Khalid and Aljamaan, Fadi and Raina, Rupesh and Malki, Khalid H and Altamimi, Ibraheem and Muaygil, Ruaim and Wahabi, Hayfaa and Jamal, Amr and Alhaboob, Ali and others},
  journal = {JMIR Medical Education},
  volume = {11},
  pages = {e63400},
  year = {2025},
  publisher = {JMIR Publications Toronto, Canada},
}

@article{peereboom2025cognitive,
  title = {Cognitive phantoms in large language models through the lens of latent variables},
  author = {Peereboom, Sanne and Schwabe, Inga and Kleinberg, Bennett},
  journal = {Computers in Human Behavior: Artificial Humans},
  pages = {100161},
  year = {2025},
  publisher = {Elsevier},
}

@article{vaswani2017attention,
  title = {Attention is all you need},
  author = {Vaswani, Ashish and Shazeer, Noam and Parmar, Niki and Uszkoreit, Jakob and Jones, Llion and Gomez, Aidan N and Kaiser, {\L}ukasz and Polosukhin, Illia},
  journal = {Advances in neural information processing systems},
  volume = {30},
  year = {2017},
}

@inproceedings{cho2025transformer,
  title = {Transformer Explainer: Interactive Learning of Text-Generative Models},
  author = {Cho, Aeree and Kim, Grace C and Karpekov, Alexander and Helbling, Alec and Wang, Zijie J and Lee, Seongmin and Hoover, Benjamin and Chau, Duen Horng Polo},
  booktitle = {Proceedings of the AAAI Conference on Artificial Intelligence},
  volume = {39},
  number = {28},
  pages = {29625--29627},
  year = {2025},
}

@book{MarchWolff1917_Calculus,
  title = {Calculus},
  author = {March, Herman W. and Wolff, Henry C.},
  publisher = {McGraw-Hill},
  year = {1917},
  address = {New York},
  pages = {xvi + 360},
  oclc = {1041616446},
  lccn = {17002039},
}

@inproceedings{Ndum2025RadiantLLM,
  title = {{RADIANT-LLM}: Retrieval-Augmented Domain Intelligent {LLM} Framework for Safe, Secure, and Safeguarded Design of Advanced Nuclear Reactor Technologies},
  author = {Ndum, Zavier Ndum and Tao, Jian and Ford, John and Mansung, Yim and Liu, Yang},
  booktitle = {Proceedings of the 66th Annual International Nuclear Materials Management (INMM) Meeting},
  year = {2025},
  month = aug,
  address = {Washington, D.C., USA},
  publisher = {Institute of Nuclear Materials Management (INMM)},
}

@article{auer2024docling,
  title = {Docling technical report},
  author = {Auer, Christoph and Lysak, Maksym and Nassar, Ahmed and Dolfi, Michele and Livathinos, Nikolaos and Vagenas, Panos and Ramis, Cesar Berrospi and Omenetti, Matteo and Lindlbauer, Fabian and Dinkla, Kasper and others},
  journal = {arXiv preprint arXiv:2408.09869},
  year = {2024},
}

@article{zhang2024document,
  title = {Document parsing unveiled: Techniques, challenges, and prospects for structured information extraction},
  author = {Zhang, Qintong and Wang, Bin and Huang, Victor Shea-Jay and Zhang, Junyuan and Wang, Zhengren and Liang, Hao and He, Conghui and Zhang, Wentao},
  journal = {arXiv preprint arXiv:2410.21169},
  year = {2024},
}

@misc{marker2025,
  title = {Marker: PDF to Markdown and JSON Document Conversion Tool},
  author = {{datalab-to/marker} developers},
  year = {2025},
  howpublished = {\url{https://github.com/datalab-to/marker}},
}

@inproceedings{papineni2002bleu,
  title = {Bleu: a method for automatic evaluation of machine translation},
  author = {Papineni, Kishore and Roukos, Salim and Ward, Todd and Zhu, Wei-Jing},
  booktitle = {Proceedings of the 40th annual meeting of the Association for Computational Linguistics},
  pages = {311--318},
  year = {2002},
}

@inproceedings{chin2004rouge,
  title = {Rouge: A package for automatic evaluation of summaries},
  author = {Chin-Yew, Lin},
  booktitle = {Proceedings of the Workshop on Text Summarization Branches Out, 2004},
  year = {2004},
}

@article{rajpurkar2016squad,
  title = {Squad: 100,000+ questions for machine comprehension of text},
  author = {Rajpurkar, Pranav and Zhang, Jian and Lopyrev, Konstantin and Liang, Percy},
  journal = {arXiv preprint arXiv:1606.05250},
  year = {2016},
}

@techreport{badwan2015application,
  title = {Application of Framework for Integrating Safety, Security and Safeguards (3Ss) into the Design Of Used Nuclear Fuel Storage Facility},
  author = {Badwan, Faris M and Demuth, Scott F},
  year = {2015},
  institution = {Los Alamos National Laboratory (LANL), Los Alamos, NM (United States)},
}

@article{yi2026gai,
  title = {GAI-HIQ: Developing a health information quality assessment indicator system for generative artificial intelligence},
  author = {Yi, Jia and Du, Fei and Nie, Yuchen and Liang, Wencui and Zhou, Xiaoyu and Chen, Jinjuan and Li, Guangying and Liu, Mimi and Lv, Yalan and Zhao, Wenlong and others},
  journal = {Information Processing \& Management},
  volume = {63},
  pages = {104651},
  year = {2026},
  publisher = {Elsevier},
}

@article{ZHENG2024110382,
title = {Empirical study on fine-tuning pre-trained large language models for fault diagnosis of complex systems},
journal = {Reliability Engineering \& System Safety},
volume = {252},
pages = {110382},
year = {2024},
author = {Shuwen Zheng and Kai Pan and Jie Liu and Yunxia Chen}
}

@article{XIAO2026112123,
title = {AutoGraph: An intelligent knowledge-graph agent for procedure automation and dynamic human reliability support in high-risk industries},
journal = {Reliability Engineering \& System Safety},
volume = {270},
pages = {112123},
year = {2026},
author = {Xingyu Xiao and Ben Qi and Zijian Yin and Jiejuan Tong and Jun Sun and Zhe Sui and Jingang Liang and Jun Zhao and Haitao Wang}
}

@article{ZHANG2026112333,
title = {A hybrid deep learning and large language models framework for ship collision accident analysis},
journal = {Reliability Engineering \& System Safety},
volume = {273},
pages = {112333},
year = {2026},
author = {Xiyu Zhang and Langxiong Gan and Hailong Cui and Yaqing Shu and Jakub Montewka and Zaili Yang}
}

@article{YU2026111973,
title = {Lightweight multimodal LLM-empowered dual-agent collaboration for reliable defect detection and maintenance recommendation in tunnel infrastructure},
journal = {Reliability Engineering \& System Safety},
volume = {268},
pages = {111973},
year = {2026},
author = {Gang Yu and Ruibin Ju and Vijayan Sugumaran and Haizhu Liu}
}

@article{LI2026112277,
title = {A deep learning framework for aviation risk classification and high-order coupled risk modeling},
journal = {Reliability Engineering \& System Safety},
volume = {271},
pages = {112277},
year = {2026},
author = {Xirui Li and Fairuz Izzuddin Romli and Syaril Azrad Md Ali and Amzari Zhahir and Junqi Tang}
}

@article{LIU2026111891,
title = {Pirate-GPT: A locally deployed large language model framework for reliable offline anti-piracy decision support and knowledge retrieval in maritime operations},
journal = {Reliability Engineering \& System Safety},
volume = {267},
pages = {111891},
year = {2026},
author = {Xiliang Liu and Jinghong Hu and Qiang Mei and Shaohua Wang}
}

@article{LIANG2026112416,
title = {Domain-specific large language model-driven risk analysis of battery energy storage systems},
journal = {Reliability Engineering \& System Safety},
volume = {274},
pages = {112416},
year = {2026},
author = {Jiali Liang and Huixing Meng and Yu Mu}
}

@article{XIAO2026111585,
title = {KRAIL: A knowledge-driven framework for human reliability analysis integrating IDHEAS-DATA and large language models},
journal = {Reliability Engineering \& System Safety},
volume = {265},
pages = {111585},
year = {2026},
issn = {0951-8320},
author = {Xingyu Xiao and Peng Chen and Ben Qi and Hongru Zhao and Jingang Liang and Jiejuan Tong and Haitao Wang}
}

@article{lewis2020retrieval,
  title={Retrieval-augmented generation for knowledge-intensive nlp tasks},
  author={Lewis, Patrick and Perez, Ethan and Piktus, Aleksandra and Petroni, Fabio and Karpukhin, Vladimir and Goyal, Naman and K{\"u}ttler, Heinrich and Lewis, Mike and Yih, Wen-tau and Rockt{\"a}schel, Tim and others},
  journal={Advances in neural information processing systems},
  volume={33},
  pages={9459--9474},
  year={2020}
}

@inproceedings{gokdemir2025hiperrag,
  title={HiPerRAG: High-performance retrieval augmented generation for scientific insights},
  author={Gokdemir, Ozan and Siebenschuh, Carlo and Brace, Alexander and Wells, Azton and Hsu, Brian and Hippe, Kyle and Setty, Priyanka and Ajith, Aswathy and Pauloski, J Gregory and Sastry, Varuni and others},
  booktitle={Proceedings of the Platform for Advanced Scientific Computing Conference},
  pages={1--13},
  year={2025}
}

@article{LIM2025100023,
title = {An AI-Driven Thermal-Fluid Testbed for Advanced Small Modular Reactors: Integration of Digital Twin and Large Language Models},
journal = {AI Thermal Fluids},
volume = {4},
pages = {100023},
year = {2025},
author = {Doyeong Lim and Zavier Ndum Ndum and Christian Young and Yassin Hassan and Yang Liu}
}

@article{ABULAWI2025111353,
title = {Bayesian Optimized Deep Ensemble for Uncertainty Quantification of Deep Neural Networks: a System Safety Case Study on Sodium Fast Reactor Thermal Stratification Modeling},
journal = {Reliability Engineering \& System Safety},
volume = {264},
pages = {111353},
year = {2025},
author = {Zaid Abulawi and Rui Hu and Prasanna Balaprakash and Yang Liu},
}

@article{LIU2021107636,
title = {Uncertainty quantification for Multiphase-CFD simulations of bubbly flows: a machine learning-based Bayesian approach supported by high-resolution experiments},
journal = {Reliability Engineering \& System Safety},
volume = {212},
pages = {107636},
year = {2021},
author = {Yang Liu and Dewei Wang and Xiaodong Sun and Yang Liu and Nam Dinh and Rui Hu}
}




\end{document}